\documentclass{aa}

\usepackage{graphicx}
\usepackage{txfonts}
\usepackage{amsmath,multirow}
\usepackage{mathtools}
\usepackage{bm}
\usepackage{natbib}
\bibpunct{(}{)}{;}{a}{}{,} 
\usepackage{hyperref}
\usepackage{float}
\usepackage{placeins}

\begin{document}

   \title{From a mass invariant in a gravo-turbulent medium to the characteristic mass of the star-forming cores}
   \titlerunning{The characteristic mass of the star-forming cores}


   \author{Pierre Dumond \inst{1}, Gilles Chabrier\inst{1,2}, J\'er\'emy Fensch \inst{1} \and Etienne Jaupart\inst{3}}

   \institute{
            ENS de Lyon, Université Lyon 1,  CNRS, CRAL, UMR 5574, Lyon, France\\
              \email{pierre.dumond@ens-lyon.fr}
         \and
             School of Physics, University of Exeter, Exeter, EX4 4QL, UK
        \and
            CEA-DAM-DIF, F-91297 Arpajon, France
            }

   \date{Received XXX; accepted XXX}

  \abstract{ The origin of the observed universality, or, say, the limited variations, of the {stellar} initial mass function (IMF) and its precursor, the  {prestellar} core mass function (CMF),
  and thus of the characteristic mass of stars and cores, across a wide range of different galactic environments remains an open, fundamental problem in astrophysics. In this paper, we demonstrate that we can explain this property from gravity and turbulence only, using a mass time-invariant that remains valid for a broad class of turbulent flows throughout the star formation process. This mass invariant is entirely determined by the large-scale normalisation of the field of density fluctuations in the Galactic Interstellar Medium (ISM), induced by turbulence. Using high-resolution numerical simulations, we first verify that the temporal invariance of this quantity is preserved in a self-gravitating turbulent medium.
  Then, introducing a self-consistent model of the statistical properties of the density field in the self-gravitating multiphase turbulent ISM, we show that the universality of the characteristic mass of star-forming cores is  imposed by this invariant together with the scaling of the Larson relations. Our model successfully predicts the characteristic mass of the cores in Milky Way-like star-forming regions, as well as in more extreme environments such as high-mass star-forming regions and massive early-type galaxies. 
  This scenario suggests that the characteristic mass of the CMF, and then probably the characteristic stellar mass, originates from a universal, fundamental property of turbulence and is imprinted both by the very early stages of the star formation process in the gravo-turbulent interstellar medium, when the dynamics of the medium are still dominated by turbulence, and by the statistical properties of small-scale self-gravitating structures. }


   \maketitle
%
\nolinenumbers

\section{Introduction}
\label{Introduction}

The star formation process owes its remarkable complexity to the interplay between multiple physical mechanisms, including self-gravity, the thermodynamic properties of the gas and magnetic fields, all operating within a highly turbulent medium over a wide range of spatial scales (see, e.g., \citealt{Offner_OriginUniversalityStellar2014, Hennebelle_PhysicalOriginStellar2024, Schinnerer_MolecularGasStarFormation2024} for reviews). Star formation involves scales spanning from at least molecular clouds of order $\sim100\,\mathrm{pc}$ down to dense cores of order $\sim1\,\mathrm{AU}$.  
As a consequence, building a comprehensive and self-consistent framework capable of explaining the observed statistical properties of star formation, notably the stellar Initial Mass Function (IMF) and the Star Formation Rate (SFR), represents a major challenge in astrophysics. Beyond their intrinsic importance for understanding the physics of star formation, such models are also of broad interest to the community, as they are key ingredients in cosmological and galactic simulations, in particular for modelling stellar feedback in star-forming regions.

One of the most striking features of the star formation process is its apparent universality across a wide range of star-forming environments, as evidenced by the near-universal nature, or, say, the  limited variations of the IMF.  The IMF and the Core Mass Function (CMF) are observed to be remarkably universal across a large variety of star-forming regions \citep{Offner_OriginUniversalityStellar2014, Konyves_CensusDenseCores2015, Hennebelle_PhysicalOriginStellar2024, Zhang_ProbingFilamentaryNature2024}. These observations suggest that the same fundamental physical mechanisms govern star formation in various typical Galactic environments.

However, despite the strong evidence for the universality of the IMF and CMF in standard Milky Way-like star-forming regions, systematic variations have also been reported for extreme star-forming environments, such as high-mass star-forming regions \citep{Motte_UnexpectedlyLargeProportion2018, Pouteau_ALMAIMFIIIInvestigating2022}, the Central Molecular Zone \citep{Hosek_UnusualInitialMass2019a} or massive early-type galaxies \citep{Smith_EvidenceInitialMass2020a,VanDokkum_ReconcilingRatiosCosmic2024}, all characterised by much higher densities and turbulence levels. In the first two examples, the high-mass slope of the IMF appears to be significantly shallower than the standard one in the MW (see \cite{Chabrier_ConsistentExplanationUnusual2024} for a detailed study) whereas in the third case the peak of the IMF is shifted towards smaller masses, indicating a larger fraction of low-mass stars ($\lesssim 1{\rm M}_\odot$) than in the MW. Therefore, understanding both this universality in MW-like standard conditions and  these deviations in extreme environments is essential for a complete understanding of star formation.

{The dependence of the IMF on environmental conditions has also been investigated through numerical simulations. In particular, \cite{Mathew_RoleTurbulenceDriving2023} showed that the nature of the large-scale turbulent forcing slightly modifies the peak of the resulting sink mass function. Similarly, \cite{Mathew_InfluenceCloudVirial2024} found that the characteristic mass of the sink distribution shifts toward lower masses as the virial parameter of the parent molecular cloud decreases.
Another parameter that may affect the IMF is the metallicity of the gas. While the high-mass slope appears to be largely insensitive to metallicity variations \citep{Tanvir_MetallicityDependenceStellar2024}, several studies have reported a dependence at the low-mass end of the IMF \citep{Li_StellarInitialMass2023, Bate_VariationLowmassEnd2025, LiLi_VariationsMilkyWays2026}. These variations, however, generally remain restricted to a relatively narrow mass range.}

Over the past few decades, several theoretical models have been proposed to explain the origin and universality of the IMF and the star formation process in MW-like conditions. These models can broadly be divided into two classes. The gravo-turbulent scenario assumes that the statistical properties of the gas are dominated by turbulence at all scales \citep{Hennebelle_AnalyticalTheoryInitial2008, Hennebelle_ANALYTICALSTARFORMATION2011, Hopkins_GeneralTheoryTurbulent2013, Guszejnov_MappingCoreMass2015}, while the global hierarchical collapse scenario supposes that the gas dynamics are governed primarily by gravity from spiral-arm scales down to the core scale \citep{Zinnecker_PREDICTIONPROTOSTELLARMASS1982, Bonnell_AccretionStellarClusters2001, Vazquez-Semadeni_GlobalHierarchicalCollapse2019c}.  
Each framework provides an explanation for the universality of the IMF, either through the universality of turbulent properties or through the universality of gravitational collapse. 
There is another fundamental difference between these two scenarios: while in the first one the properties of the IMF are determined at the origin of the star formation process and are imprinted in the large-scale properties of the parent cloud (mean density, large-scale turbulent velocity, temperature), in the second scenario the IMF is determined essentially at the final stage of the collapse, during the gas-to-star conversion process. However, we now know that none of these two scenarios is entirely correct, since the statistical properties and dynamical state of the gas density and velocity fluctuations in star-forming gravo-turbulent molecular clouds are set by an { interplay} between at least gravity and turbulence, and potentially the magnetic field \citep{Jaupart_EvolutionDensityPDF2020}.
The physical origin of the observed universality of the star formation process thus remains an open and actively debated question.

One of the major challenges faced by these theoretical frameworks is to account for the extremely broad range of involved spatial scales  and to describe consistently the temporal evolution of gravitational collapse in the cloud. To address these issues, we rely in this paper on a temporal invariant derived by \cite{Chandrasekhar_FluctuationsDensityIsotropic1951} from the mass conservation equation in a turbulent medium. This invariant holds for ergodic\footnote{Ergodicity and thus statistical (not to be confused with spatial) homogeneity are the necessary underlying assumptions of any study aiming at giving a { global} description, either observational or theoretical, of structure statistical properties in astrophysics and cosmology \citep{Peebles_StatisticalAnalysisCatalogs1973, Frisch_TurbulenceLegacyKolmogorov1995}.} (thus statistically homogeneous) and isotropic flows with a steady-state background, provided that the cross-correlation function $C_{\rho,\rho\bm{v}}(|\bm{q}|)$ between the density field $\rho$ and the momentum field $\rho\bm{v}$ decays faster than $1/|\bm{q}|^2$ at large separations, where $\bm{q}$ denotes the spatial separation.

Recently, \citet{Jaupart_GeneralizedTransportEquation2021a} extended this invariant to non-isotropic turbulence and to flows with a time-evolving statistical mean density $\mathbb{E}(\rho)(t)={\bar \rho}(t)$, thus describing a possible global shear, compression or expansion, obtaining the following expression:
\begin{equation}
    \label{eq_M_inv_def}
    M_{\rm inv} = \mathbb{E}(\rho)(t)\,\mathrm{Var}\!\left(\frac{\rho}{\mathbb{E}(\rho)}\right)_t (l_{\rm c}^\rho)_t^3 = \mathrm{const},
\end{equation}
where $l_{\rm c}^\rho$ is the correlation length of the density field, defined as
\begin{equation}
    \label{eq:lc}
    l_{\rm c}^\rho = \left(\frac{1}{8C_\rho(0)}\int_{\mathbb{R}^3} C_\rho(\bm{q})\,{\rm d}^3\bm{q}\right)^{1/3},
\end{equation}
and $C_\rho$ denotes the autocovariance function (ACF) of the density field. {If the mean density of the medium is constant, an equivalent formulation of this invariant is the value at zero-wavenumber of the power spectrum of the density field \footnote{The power spectrum at zero wavenumber is not the squared mean density of the field (and is not related by any means to it) if the power spectrum is properly defined as the Fourier transform of the autocovariance function.}.}
The validity of this invariant has already been verified in decaying turbulent flows by \citet{Dumond_MassInvariantCompressible2025}.

Building on the aforementioned invariant, we investigate in this paper the universality of the star formation process through a study of the characteristic mass of dense star-forming cores. We show that the universality of the characteristic core mass can be explained by this temporal invariant that is preserved throughout the star formation process in molecular clouds obeying the \cite{Larson_TurbulenceStarFormation1981}  density and velocity scaling relations. We first present a numerical verification of the time invariance of $M_{\rm inv}$ in a self-gravitating medium. We then propose a model for computing the characteristic core mass based on this invariant and demonstrate that it provides a robust explanation for the observed universality of the mean core mass in the CMF. Finally, we discuss the implications of our results and compare our predictions to observations in various star formation contexts.

\section{Numerical verification of the invariant in self-gravitating medium}

\subsection{Numerical setup}

The numerical setup used in the present study is essentially identical to that employed in \citet{Dumond_MassInvariantCompressible2025}. Its main features are summarised below.
We perform fixed-grid simulations using the hydrodynamical code \textsc{RAMSES}. The numerical scheme is based on a second-order Godunov solver, and turbulence is driven through an Ornstein-Uhlenbeck process applied to the acceleration field in Fourier space. The originality of these simulations is that turbulence is forced at scales significantly smaller than the box size, namely at $L_{\rm box}/7$. As shown in \cite{Dumond_MassInvariantCompressible2025}, this choice ensures that statistical estimates of the correlation length, the density variance, and the invariant $M_{\rm inv}$ are robust. Periodic boundary conditions are adopted. The equation of state is isothermal with a temperature \(T=10\,{\rm K}\) corresponding to a sound speed of \(c_{\rm s}=220\,{\rm m\,s}^{-1}\).

To test the temporal invariance of $M_{\rm inv}$ (see Eq.~\ref{eq_M_inv_def}) in a self-gravitating medium, we first drive the turbulence for at least two turbulent crossing times, $\tau_{\rm cross}$, before activating gravity. This ensures that the turbulence is fully developed when gravity is switched on \citep{Federrath_UniversalitySupersonicTurbulence2013}. The external driving remains switched on even after self-gravity has been activated. In the present study, we only consider the case of a constant mean density \(\mathbb{E}(\rho)\).

We perform simulations at two different large-scale Mach numbers, $\mathcal{M}=5$ and $\mathcal{M}=10$, in order to test the invariant under different dynamical conditions. The box size \(L_{\rm box}\), the mean number density \(\bar{n}\) and the 3D velocity dispersion are chosen to approximately satisfy the Larson relations \citep{Larson_TurbulenceStarFormation1981, Hennebelle_TurbulentMolecularClouds2012,Miville-Deschenes_PHYSICALPROPERTIESMOLECULAR2017}:
\begin{align}
    \label{eq_larson_velocity}
    \sigma_{3{\rm D}} & \simeq V_0 \left(\frac{L_{\rm box}}{1~\mathrm{pc}}\right)^{0.5}, \\
    \label{eq_larson_density}
    \bar{n} & \simeq d_0 \left(\frac{L_{\rm box}}{1~\mathrm{pc}}\right)^{-1}.
\end{align}
with \(V_0=600~\mathrm{m~s^{-1}}\) and \(d_0 = 800~\mathrm{cm^{-3}}\). 
We also consider several spatial resolutions to assess the convergence of our results, as shown in Appendix \ref{App_convergence}. We also choose values for the Mach number and the mean density of the box higher than those that would be prescribed by the above relations in order to accelerate the dynamics and save computational time, but keeping virial parameters representative of molecular clouds. The virial parameter \(\alpha_{\rm vir}\simeq(\tau_{\rm ff}/\tau_{\rm cross})^2\) ranges between 2 and 4, where \(\tau_{\rm ff}\) is the free-fall time-scale at the mean density of the medium.
A summary of the simulations is provided in Table~\ref{tab_simus_list}.

\begin{table*}
    \begin{center}
    \caption{List of simulations.}
    \begin{tabular}{ccccccccc}
        \(\mathcal{M}\) & \(\bar{n} \ ({\rm cm}^{-3})\) & \(L_{\rm box}\) (pc) &  \(N_{\rm cells}\) & $\sigma_s^2$ & \(\tau_{\rm cross}\) (Myr) & \(\tau_{\rm ff}\) (Myr) & \(\alpha_{\rm vir}\) & Fiducial\\
        \hline \multirow{2}{*}{5} & \multirow{2}{*}{\(4.9\times 10^3\)} & \multirow{2}{*}{ \(3.5\)} &  \(1024^3\) & 2.94 & \multirow{2}{*}{ \(0.4\)} & \multirow{2}{*}{\(0.9\)} & \multirow{2}{*}{\(4\)} & \\
        & & & \(2048^3\) & 2.83 & & & & *\\
        \hline \multirow{2}{*}{10} & \multirow{2}{*}{\(1.8\times 10^3\) } & \multirow{2}{*}{\(14\)} &  \(1024^3\) & 5.44 & \multirow{2}{*}{\(0.9\)} & \multirow{2}{*}{ \(1.5\)} & \multirow{2}{*}{\(2.8\)} & \\
        & & & \(2048^3\) & 4.95 & & & & *\\
    \end{tabular}
    \end{center}
     \tablefoot{From left to right: the 3D rms Mach number \(\mathcal{M}\), the mean number density \(\bar{n}\), the size of the box \(L_{\rm box}\), the number of cells \(N_{\rm cells}\), the variance of the log-density field \(\sigma_s^2\) (\(s=\rho/{\bar \rho}\)) before the activation of gravity, the turbulent crossing time \(\tau_{\rm cross}=L_{\rm box}/(\mathcal{M}c_{\rm s})\), the free-fall timescale \(\tau_{\rm ff} = 1/\sqrt{\bar{\rho}G}\) and the virial parameter \(\alpha_{\rm vir}\simeq(\tau_{\rm ff}/\tau_{\rm cross})^2\). The sign \(*\) denotes the fiducial runs used in this study.}
    \label{tab_simus_list}
\end{table*}

\subsection{Numerical computation of the invariant}

Following \citet{Dumond_MassInvariantCompressible2025}, the correlation length (Eq.~\ref{eq:lc}), the variance, and the invariant (Eq.~\ref{eq_M_inv_def}) are not computed over the full periodic domain but rather within a sub-box of size $L_{\rm box}/2$. This approach avoids the presence of spurious large-scale correlations between $\rho$ and $\rho\bm{v}$ that arise due to the periodicity of the entire simulated volume. The rapid decay of this cross-correlation at large separations is indeed a key assumption for the temporal invariance of $M_{\rm inv}$. The correlation length is computed by truncating the integration of the autocovariance function at its first zero-crossing. At larger separations, the autocovariance is dominated by noise, owing to the absence of statistical information at scales exceeding the sub-box size. During the simulations, the evolution of the invariant is tracked as long as the correlation length is resolved by at least ten grid cells. The robustness of this method for estimating the correlation length has been demonstrated in \citet{Dumond_MassInvariantCompressible2025}.

\subsection{Verification of the invariant}

\begin{figure*}
    \centering
    \includegraphics[width=\textwidth]{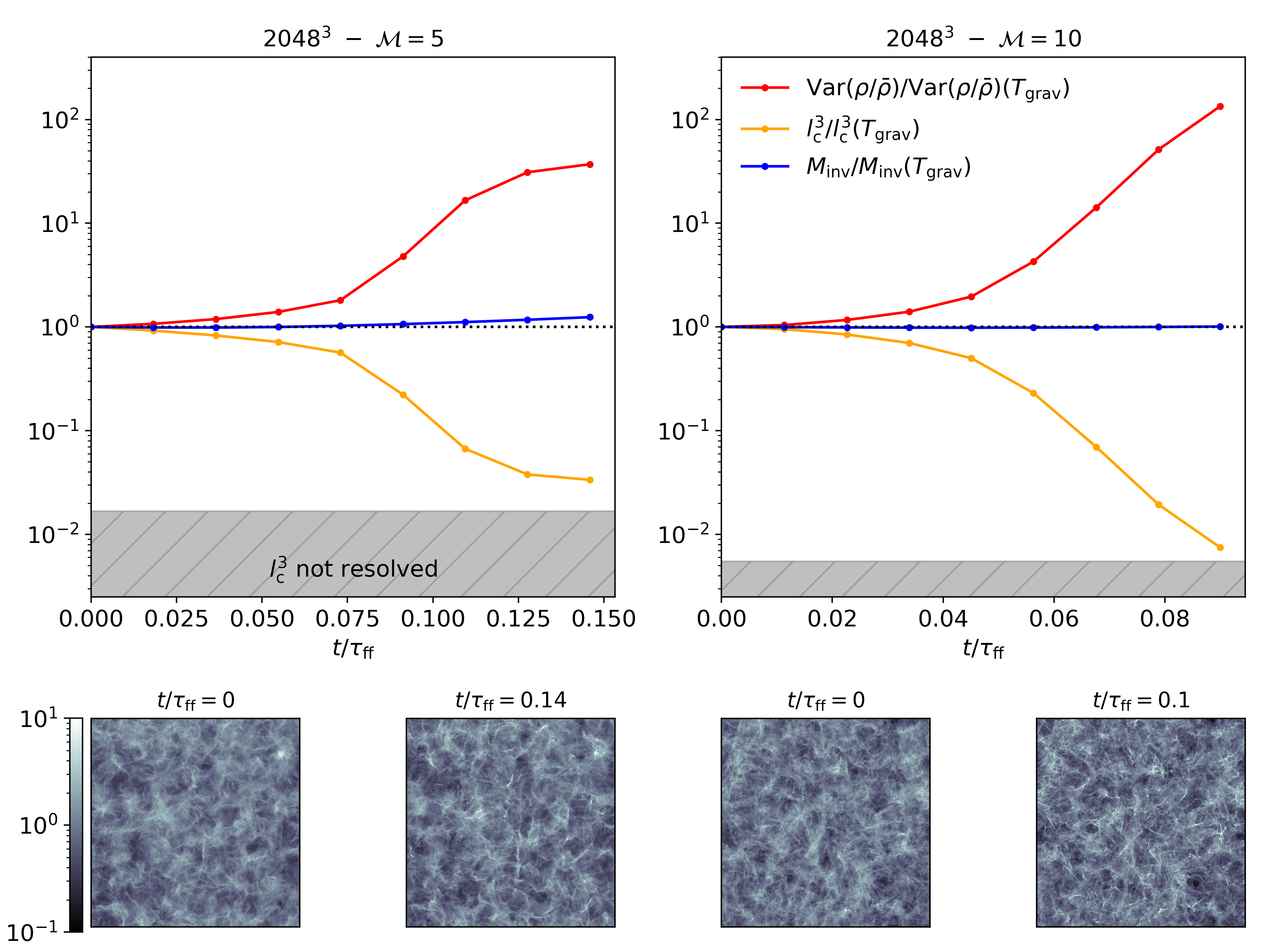}
    \caption{Evolution of mass \(M_{\rm inv}\) (blue), correlation length \(l_{\rm c}^3\) (orange) and density variance \(\mathrm{Var}(\rho/\mathbb{E}(\rho))\) (red) after the activation of gravity at \(t=0\) for two Mach numbers (and thus two corresponding sizes and mean densities following Eqs. \ref{eq_larson_density} and \ref{eq_larson_velocity}) in the 2048\(^3\) simulations (see Table~\ref{tab_simus_list}). The integration is stopped when the correlation length is no longer well resolved. All three quantities shown are normalised to their value at the time at which self-gravity is switched on. The time is normalised to the large-scale free-fall time \(\tau_{\rm ff} = \frac{1}{\sqrt{G\bar{\rho}}}\).
    The four bottom panels show the evolution of the projected density field from the initial state without self-gravity and the final state, before the correlation length is no longer resolved.}
    \label{fig_M_inv_grav}
 \end{figure*}

In Fig.~\ref{fig_M_inv_grav}, we show the temporal evolution of the invariant from the time at which gravity is activated in the simulation until the correlation length is no longer resolved by at least ten grid cells. We find that the temporal invariance of $M_{\rm inv}$ is well satisfied, even though the density variance increases by approximately two orders of magnitude during gravitational collapse. A departure of $\sim30\%$ observed in the invariant at \(\mathcal{M}=5\) is attributed to limited statistical sampling due to the finite simulation volume. 
The limitation is lower at higher Mach numbers since the correlation length decreases as the Mach number increases. This increases the accuracy of statistical estimates of variance and correlation length. 
This highlights the intrinsic difficulty of numerically verifying such an invariant, which requires both sufficiently high spatial resolution to resolve the decreasing correlation length and a sufficiently large statistical volume to obtain accurate estimates of the relevant statistical quantities, namely the variance and the correlation length. {Since the mean density of the medium remains constant throughout our numerical simulations, we validated the equivalent formulation of the invariant using the zero-wavenumber value of the density field power spectrum. As demonstrated in Appendix~\ref{App_invariant_num_PS}, the value of the density power spectrum at the largest accessible mode exhibits negligible variation during gravitational collapse.}

In the present study, we restrict ourselves to an isothermal equation of state. However, in \citet{Dumond_MassInvariantCompressible2025}, we have already shown that the invariant remains constant in decaying turbulent flows for different equations of state, including both isothermal and adiabatic cases. More generally, the equation of state, or any other local physical process, is not expected to violate the invariance of $M_{\rm inv}$. Indeed, mechanisms that would break the invariance should create a cross-correlation between the density and the momentum that does not decay sufficiently rapidly at large separations. Such behaviour is unlikely to arise from small-scale local processes such as thermodynamics. Consequently, even in media characterised by complex thermochemical properties, such as the interstellar medium, we expect the invariant to remain preserved and to be insensitive to heating and cooling processes.

We emphasise that the invariant considered here is not related to the total mass of the simulated turbulent gas. As shown by \citet{Dumond_MassInvariantCompressible2025}, the value of the invariant $M_{\rm inv}$ is instead linked to the injection mass, $\bar{\rho} L_{\rm inj}^3$, where $\bar{\rho}$ is the mean gas density and \(L_{\rm inj}\) the injection length, and not to the total mass contained in the computational domain. More fundamentally, the invariant is not associated with the mass of the system itself, as it remains well defined even when the domain of interest is $\mathbb{R}^3$ if we keep a finite injection length. In this limit, $M_{\rm inv}$ remains finite whereas the total mass of the system diverges. 

In the following section, we demonstrate how the temporal invariance of $M_{\rm inv}$ can be used to predict the characteristic core mass across different star-forming environments and to discuss its universality.

\section{Astrophysical application: from the mass of the invariant to the characteristic mass of the cores}
\label{Sec_Astro_application}

In the previous section, we numerically demonstrated that the quantity $M_{\rm inv}$ remains constant in a highly non-stationary self-gravitating medium. Applied to the context of star formation, this result suggests that $M_{\rm inv}$ is conserved throughout the entire star formation process, from the initial conditions corresponding to the injection of turbulence into the interstellar medium at large scales to the final stage of stellar birth at core scales.

As already pointed out by \citet{Jaupart_GeneralizedTransportEquation2021a}, this invariance may imply that a characteristic mass is imprinted at the very early stages of star formation, namely when turbulence is injected into the medium, leading to random over- and under-density fluctuations, and is subsequently preserved during the collapse. In their work, the authors directly associate the invariant mass with the mass of small-scale structures, i.e. the dense cores formed at the end of the collapse.

However, as shown by \citet{Chandrasekhar_FluctuationsDensityIsotropic1951}, the invariant is primarily related to the value of the density power spectrum at zero wavenumber. This implies that it is fundamentally connected to large-scale density structures rather than to small-scale ones. While the invariant does impose constraints on small-scale structure formation through the density power spectrum, it is not possible to establish a direct equality between the invariant mass and the mass of individual collapsing objects.

In order to relate the invariant mass to the mass of dense cores formed during gravitational collapse in the interstellar medium, we introduce a statistical model of the density field in the self-gravitating ISM that consistently incorporates both its power spectrum and its probability density function (PDF). This model and its predictions are presented in the following sections.

\subsection{Qualitative scenario of the initial conditions of star formation}

\begin{figure}
    \centering
    \includegraphics[width=\columnwidth]{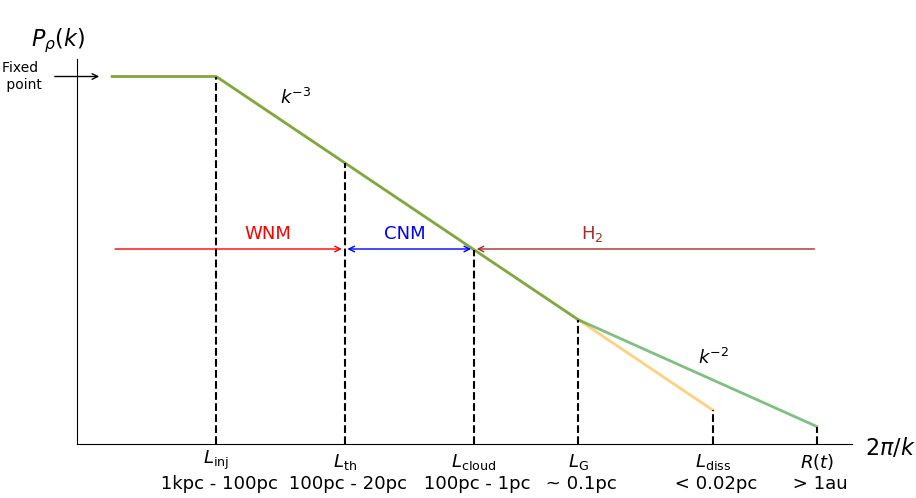}
    \caption{Schematic representation of the star formation process through the evolution of the 3D density power spectrum. The yellow spectrum corresponds to the (hypothetical) initial conditions before gravity significantly affects the dynamics of the medium. The green spectrum describes the two-point statistics of the gas during the collapse process, from the injection scale $L_{\rm inj}$ down to the characteristic core size $R(t)$. As collapse proceeds, $R(t)$ decreases until it reaches the first Larson core scale (1~AU).}
    \label{Fig_schema_scenario}
\end{figure}

 The star formation scenario we rely on, which will guide the following derivations, is outlined below  and developed in the next subsections. A schematic representation of this framework, illustrated through the functional form of the density power spectrum from the injection scale down to the core scale, is shown in Fig.~\ref{Fig_schema_scenario}.

 We consider turbulence to be injected within spiral arms, typically on scales ranging from $\sim100\,\mathrm{pc}$ to $\sim1\,\mathrm{kpc}$ in a medium of mean density \(\bar{\rho}\). The dominant physical mechanisms responsible for driving this turbulence remain uncertain; the most plausible candidates, which are not mutually exclusive, are supernova feedback and gravitational instabilities \citep{MacLow_ControlStarFormation2004, Bacchini_EvidenceSupernovaFeedback2020, Colman_SignatureLargeScale2022}. The injected turbulence subsequently cascades toward smaller scales, generating density fluctuations in the warm neutral medium (WNM) and cold neutral medium (CNM), and ultimately leading to the formation of molecular clouds, which are the sites of star formation. Within these molecular clouds, turbulence becomes increasingly influenced by the self-gravity of the gas which acts as a kind of `catalytic process', modifying the density field and driving density fluctuations into smaller and smaller, denser and denser structures. 
 At small scales, i.e. typically below the scale of the formed filaments, the gas thus undergoes an approximately isothermal collapse until it reaches the first Larson core scale \citep{Larson_NumericalCalculationsDynamics1969,Masunaga_RadiationHydrodynamicModel1998, Vaytet_SimulationsProtostellarCollapse2012}.
 
 Throughout this entire process, the invariant $M_{\rm inv}$ that characterises the density fluctuations formed by the injected energy is conserved, implying that the value of the density power spectrum at the injection scale remains fixed. 
 As a consequence, given a functional form for the density power spectrum from the injection scale down to the characteristic scale of collapsing cores, together with the variance of the density field, we show in the following that it is possible to predict the characteristic mass of the prestellar cores. 

To compute the characteristic core mass, we first define the reference structure considered in this work (Sect.~\ref{Sec_collapse}): a Bonnor--Ebert sphere whose central plateau of density $\rho_{\rm max}$, surrounded by a collapsing envelope, increases as collapse proceeds, while the outer density,
{i.e. the density at the outer edge of the envelope}, remains constant. This outer density corresponds to the typical density threshold above which the gas becomes self-gravitating.
Determining the characteristic mass {at a time $t$} of such a core therefore requires tracking the evolution of both its inner radius $R$ and its maximum density $\rho_{\rm max}$ during collapse. To this end, we compute the density variance using two complementary approaches: one based on the density power spectrum (Sect. \ref{Sec_density_PS}) and the other one on the density PDF (Sect. \ref{Sec_density_Var_PDF}). These two approaches respectively link $R$ and $\rho_{\rm max}$ to the density variance.
To relate $R$ to the variance, we introduce a model for the density power spectrum, summarised in Fig.~\ref{Fig_schema_scenario}. To relate $\rho_{\rm max}$ to the variance, since no complete description of the PDF of the multiphase ISM currently exists, we first estimate the molecular gas mass fraction using the functional form of the density power spectrum corresponding to a hypothetical turbulence-dominated ISM, prior to the influence of gravity. We then compute the variance of the total ISM using the theory of the density PDF in self-gravitating media developed by \citet{Jaupart_EvolutionDensityPDF2020}, which determines the onset of the two power-law tails in the high density domain of the PDF induced by gravity\footnote{As shown in \citet{Jaupart_EvolutionDensityPDF2020}, the first power law occurs at the very early stages of the star formation process and does not last very long. Very quickly, the second power law occurs, corresponding to the density above which the gas is entirely dominated by gravity, yielding the formation of the cores, the process of interest in the present context. Throughout this paper, we thus consider only the second power-law tail.}.
The value of the characteristic core mass, as well as its dependence on the size of the parent Larson-type molecular clouds, is then discussed. The main notations used
in the article are presented in Table \ref{tab_notations}.

\begin{table*}
    \begin{center}
    \caption{Main notations used in the article.}
    \label{tab_notations}
    \begin{tabular}{cc}
       \hline Notation & Description \\
        \hline \(G\) & Gravitational constant \\
        $c_{\rm s}$ & Sound speed \\
        \(\bar{\rho}\) & Mean density of the interstellar medium at the galactic scale \\
        \(\bar{\rho}_{\rm cloud}\) & Mean density of the molecular cloud \\
        \(L_{\rm inj}\) & Injection length of turbulence \\
        \(L_{\rm th}\) & Cooling length that sets the transition between WNM and CNM \\
        \(L_{\rm cloud}\) & Diameter of the large-scale molecular cloud\\
        \(L_{\rm diss}\) & Dissipation length of turbulence \\
        \hline \(\mathcal{M}\) & 3D Mach number in the molecular cloud of size \(L_{\rm cloud}\) \\
        \(b\) & Forcing parameter of the turbulent flow \\
        \(d_0\) & Normalisation at 1~pc of the density Larson relation \\
        \(V_0\) & Normalisation at 1~pc of the 3D velocity dispersion Larson relation \\
        \(\sigma_{\rm cloud}^2\) & Variance of the density field in the cloud \\
        \hline 
        \(M_{\rm inv}\) & Invariant mass \\
        \(l_{\rm c}^\rho\) & Correlation length of the density field \\
        \(\alpha\) & Invariant mass normalised to the injection mass \(\bar{\rho}L_{\rm inj}^3\) \\
        \hline \(\rho_{\rm max}(t)\) & Maximum density of a typical collapsing core at a given time \(t\) \\
        \(R(t)\) & Inner radius of a typical collapsing core at a given time \(t\) \\
        \(\kappa\) & Slope of the envelope of the density profile of the collapsing cores \\
        \hline \(\rho_{\rm G}\) & Typical density threshold above which gravity dominates the gas dynamics\\ 
        \(L_{\rm G}\) & Typical scale threshold below which gravity dominates the gas dynamics\\ 
        \(\epsilon\) & Mass fraction of collapsing gas \\
        \(\epsilon_{\rm cloud}\) & Mass fraction of molecular clouds \\
        \(\epsilon_{\rm G}\) & Mass fraction of collapsing gas within the molecular clouds \\
       \hline
    \end{tabular}
    \end{center}
    \end{table*}

The present model is certainly a simplified description of the star formation process, as it accounts only for turbulence and gravity. In particular, we neglect the effects of magnetic fields, outflows and dynamical interactions between the cores. While these processes play some role in star formation, we assume that they are not the dominant processes responsible for setting up the characteristic mass of prestellar cores. Comparison between our results and observations will allow us to verify this assumption.

\subsection{Model for collapsing structures}
\label{Sec_collapse}

The collapsing structures considered in this work are assumed to evolve isothermally and are therefore modelled as Bonnor--Ebert-type cores, characterised by the following radial density profile:
\begin{equation}
    \label{eq_radial_profile}
    \rho(r) =
    \begin{cases}
        \rho_{\rm max}, & r < R, \\
        \rho_{\rm max}\left(\dfrac{r}{R}\right)^{-\kappa}, & R < r < R_{\rm out},
    \end{cases}
\end{equation}
where $R$ denotes the inner flat radius of the core, $\rho_{\rm max}$ its central (maximum) density at a given stage of the collapse, $\kappa$ the slope of the outer density profile, and $R_{\rm out}$ the outer radius of the core.

This functional form corresponds to a simplified version of the Plummer-like profiles commonly inferred from both observations \citep{Ward-Thompson_InitialConditionsIsolated1999, Palau_DoesMagneticField2021, Gieser_PhysicalChemicalStructure2021} and numerical simulations \citep{Masunaga_RadiationHydrodynamicModel1998, Ahmad_BirthEarlyEvolution2023}. In these studies, the slope of the outer profile is consistently found to be close to $\kappa = 2$, in agreement with theoretical predictions \citep{Penston_DynamicsSelfgravitatingGaseous1969, Larson_NumericalCalculationsDynamics1969, Gomez_DensityProfileEvolution2021, Donkov_EvolutionMassDensity2025}. We therefore adopt $\kappa = 2$ in the following.

We assume that the typical density above which structures in the cloud become self-gravitating and thus exhibit the radial profile given by Eq.~\ref{eq_radial_profile} is defined by the threshold density $\rho_{\rm G}$ at which the density PDF develops a power-law tail, $\mathbb{P}(\rho) \propto \rho^{-5/2}$ \citep{Jaupart_EvolutionDensityPDF2020}. Regions with densities above this threshold are statistically dominated by local self-gravity. This threshold density is given by \citep{Jaupart_EvolutionDensityPDF2020}:
\begin{equation}
    \label{eq_rho_sg}
    \frac{\rho_{\rm G}}{\bar{\rho}_{\rm cloud}} - 1
    = b^2 \mathcal{M}^4
    \frac{5 c_{\rm s}^2}{\pi G L_{\rm cloud}^2 \bar{\rho}_{\rm cloud}}
    \frac{\ln(\rho_{\rm G}/\bar{\rho}_{\rm cloud}) + \sigma_s^2/2}{\sigma_s^2},
\end{equation}
where $\mathcal{M}$ is the Mach number of the self-gravitating isothermal molecular cloud of size $L_{\rm cloud}$ and mean density $\bar{\rho}_{\rm cloud}$. The quantity $\sigma_s^2 = \ln[1 + (b\mathcal{M})^2]$ denotes the variance of the logarithmic density field, with $b$ the turbulent forcing parameter, typically ranging between 0.3 and 1 \citep{Federrath_ComparingStatisticsInterstellar2010}.

With this description, the characteristic mass of self-gravitating cores is given by
\begin{align}
    M_{\rm core}\simeq 4\pi R^3 \rho_{\rm max}
    \left(\frac{\rho_{\rm max}}{\rho_{\rm G}}\right)^{1/2},
    \label{eq_mass_struc_BE}
\end{align}
considering that \(R_{\rm out}\gg R\) (or equivalently \(\rho_{\rm max}\gg \rho_{\rm G}\)). We assume that $R_{\rm out}$ is defined by the condition $\rho(R_{\rm out}) = \rho_{\rm G}$, implying that outside the collapsing core, the dynamics {of the flow are mainly governed by turbulence, not by gravity} and the density therefore remains below $\rho_{\rm G}$.  

In the present description of collapsing cores, we neglect accretion flows from the turbulence-dominated regime ($\rho < \rho_{\rm G}$) onto the gravity-dominated cores ($\rho > \rho_{\rm G}$) {once the collapse has finished, i.e. \(R_{\rm out}\gg R\)}. This assumption is consistent with the physical interpretation of $\rho_{\rm G}$ introduced by \citet{Jaupart_EvolutionDensityPDF2020} within the transport equation of the mass flux, namely the density at which the mass flux driven by turbulence and gravity from turbulence-dominated regions toward gravity-dominated regions balances the mass flux driven by turbulence in the opposite direction.

{At this stage, it is already possible to derive a first estimate of the characteristic core mass \(M_{\rm core}\). At the onset of collapse, we have \(\rho_{\rm max}\simeq \rho_{\rm G}\). As shown in Appendix \ref{App_unviserval_L_grav}, for virialised clouds with \(\alpha_{\rm vir}=1\), \(\rho_{\rm G}\simeq \rho_{\rm cloud}(b\mathcal{M})^2\), which corresponds to the post-shock density. Physically, this means that gravity begins to dominate {the formation of density fluctuations} after a single shock.
As the post-shock Mach number is below 1 in isothermal turbulence due to mass conservation (shown by the Rankine-Hugoniot jump conditions), we assume that \(R_{\rm out}\) is given by the Jeans length at density \(\rho_{\rm G}\), namely
\(
R_{\rm out}\sim \frac{c_{\rm s}}{\sqrt{G\rho_{\rm G}}}.
\)
Using the Larson relations (Eqs.~\ref{eq_larson_density} and \ref{eq_larson_velocity}), this density does not depend on the cloud size and can be written as \(\rho_{\rm G}\sim d_0 b^2\mathcal{M}_0^2\) where \(\mathcal{M}_0=V_0/c_{\rm s}\) and \(d_0\) is the 1~pc normalisation of the large-scale Mach number and mean density in molecular clouds.

Neglecting geometrical factors for simplicity, we obtain
\begin{align}
M_{\rm core} &\sim \rho_{\rm G}R_{\rm out}^3 
\sim \frac{c_{\rm s}^4}{\sqrt{d_0 G^3(bV_0)^2}} \\
&\sim 3\,{\rm M}_\odot 
\left(\frac{T}{10\,{\rm K}}\right)^2
\left(\frac{d_0}{800\,{\rm cm}^{-3}}\right)^{-1/2}
\left(\frac{bV_0}{0.75\times 600\,{\rm m\,s^{-1}}}\right)^{-1}.
\end{align}
This is basically the same expression as the Bonnor-Ebert mass, but taking the post-shock density rather than the mean density of the star-forming cloud. 
This estimate is close to the observed mean core mass, typically between \(0.5\) and \(2\,{\rm M}_\odot\) (see Sects.~\ref{Sec_CoreMass} and \ref{Sec_discussion_mean_mass} for references and discussion). Importantly, this value does not depend on the size of the star-forming molecular clouds and is therefore universal once the Larson normalisations and temperature are fixed, which is likely the case for a given class of star-forming regions. Variations across star-forming environments will be discussed in Sect.~\ref{Sec_Discussion_Star_forming_env}.

This estimate of the characteristic core mass relies on zeroth-order assumptions, namely that gravity begins to dominate the dynamics after a single shock and that the outer radius of the typical core corresponds to the Jeans length at the density \(\rho_{\rm G}\). To go beyond these assumptions and obtain a more accurate and detailed estimate of the characteristic core mass, we will relate the evolution of \(R\) and \(\rho_{\rm max}\) during collapse using the mass invariant \(M_{\rm inv}\) together with the statistical properties of the gravo-turbulent medium.}

\subsection{The density power spectrum of collapsing structures}
\label{Sec_density_PS}

In analogy with the density PDF, which can be divided into two regimes, a turbulence-dominated regime at low densities and a gravity-dominated regime at high densities, the density power spectrum is also expected to exhibit two distinct regimes separated by a characteristic scale $L_{\rm G}$, below which gravity becomes dominant. On large scales, between the injection scale $L_{\rm inj}$ and $L_{\rm G}$, which typically corresponds to molecular cloud scales, the power spectrum is expected to follow a slope close to $k^{-3}$, based on the results of \cite{Rastegar_ExactTwopointCorrelation1998}, \cite{Dumond_SlopePowerSpectrum2026} and on numerous observations of turbulence-dominated regions of the interstellar medium \citep{Miville-Deschenes_ProbingInterstellarTurbulence2016a, Pingel_MultiphaseTurbulenceDensity2018, Pineda_ProbingPhysicsStarFormation2024a}. Significant variations of this slope have been reported in numerical simulations, typically ranging from $k^{-2}$ to $k^{-3.5}$ \citep{Kim_DensityPowerSpectrum2005, Federrath_STARFORMATIONEFFICIENCY2013, Konstandin_MachNumberStudy2016, Brucy_InefficientStarFormation2024a}. However, as shown by \cite{Dumond_SlopePowerSpectrum2026}, the slope of the density power spectrum measured in such simulations depends on the ratio between the injection and dissipation scales and therefore indirectly on the Reynolds number and the numerical resolution of the simulation. Since both this scale separation and the Reynolds number are necessarily much smaller in numerical simulations than in the interstellar medium, the slopes measured in simulations are not expected to accurately represent the density power spectrum of turbulence in the ISM.

Within the turbulence-dominated regime, we further assume that the power spectrum can be divided into two subdomains: one associated with the neutral gas on the largest scales and another associated with the molecular clouds on smaller scales. The transition between these two phases occurs at a characteristic scale $L_{\rm cloud}$. We assume that turbulence in both the neutral gas (WNM+CNM) and in the molecular gas follows the same cascade, consistent with several observational studies \citep{Martin_GHIGLSMappingIntermediate2015, Marchal_ROHSARegularizedOptimization2019, Marchal_ThermalTurbulentProperties2021, Elmegreen_PowerSpectraJWST2025}, although \citet{Choudhuri_TurbulentPowerSpectrum2019} reported a surprisingly steep slope in the CNM component. While this simplified description of the multiphase interstellar medium does not capture its full complexity, it provides a useful first-order framework.

In gravity-dominated regions, collapsing structures typically exhibit density profiles of the form $\rho(r)\propto r^{-2}$, which implies a density power spectrum slope proportional to $k^{-2}$. Indeed, as shown in Appendix~\ref{App_PS_grav}, a collapsing structure with a 3D density profile $\rho(r)\propto r^{-\kappa}$ exhibits a density power spectrum
\begin{equation}
    P(k) \propto k^{-6+2\kappa},
\end{equation}
for \(\kappa\in]1.5, 2.5[\). For $\kappa=2$, this yields $P(k)\propto k^{-2}$, a slope that appears to be widely observed in actively star-forming regions (see Table~3 of \citealt{Federrath_STARFORMATIONEFFICIENCY2013}).

We further assume that the transition scale below which gravity-dominated structures will start collapsing is the Jeans scale evaluated at the above-defined critical density $\rho_{\rm G}$, i.e.
\begin{equation}
    \label{Eq_L_sg}
    L_{\rm G} = \frac{\sqrt{\pi} c_{\rm s}}{\sqrt{G\rho_{\rm G}}}.
\end{equation}
Whereas, as mentioned above, the threshold density $\rho_{\rm G}$ for collapsing density regions is to be understood in a statistical sense \citep{Jaupart_EvolutionDensityPDF2020}, the scale $L_{\rm G}$ characterises the typical size of such a small-scale structure. 
{As discussed above, turbulence is not expected to contribute significantly to the support of the structures under consideration, given that they are formed through shocks. According to the Rankine-Hugoniot relation, the turbulence in the post-shock region, where the density fluctuations of interest reside, is subsonic. Consequently, the relevant characteristic scale is the sonic Jeans length rather than the turbulent Jeans length \citep{Hennebelle_AnalyticalTheoryInitial2008, Hopkins_GeneralTheoryTurbulent2013}. Furthermore, although the Jeans length formally corresponds to the smallest unstable wavelength in a homogeneous medium, we interpret it here as the largest scale at which gravity dominates over other forces. This interpretation in the present context is motivated by the highly inhomogeneous nature of the medium, in which turbulence acts as an effective filter against gravitational collapse in the pre-shock regions, i.e., outside the density fluctuations \citep{Chabrier_DimensionalArgumentImpact2011a}.}
We then note that an interesting consequence of the standard density and velocity Larson scaling relations is that the value of $\rho_{\rm G}$, solution of Eq. \ref{eq_rho_sg}, depends very weakly on the cloud scale and is found to be $\rho_{\rm G}\simeq 2\times 10^4\,{\rm cm}^{-3}$ for cloud sizes ranging from $\sim 1$ to 100 pc. 
As shown in Appendix~\ref{App_unviserval_L_grav}, this universality is a direct consequence of the Larson scaling relations. 
This value is close to the star formation threshold suggested by \citet{Lada_STARFORMATIONRATES2010}. However, the density \(\rho_{\rm G}\) computed here should be understood as a characteristic density of star formation rather than a star formation threshold.

\begin{figure*}
    \centering
    \sidecaption
    \includegraphics[width=12cm]{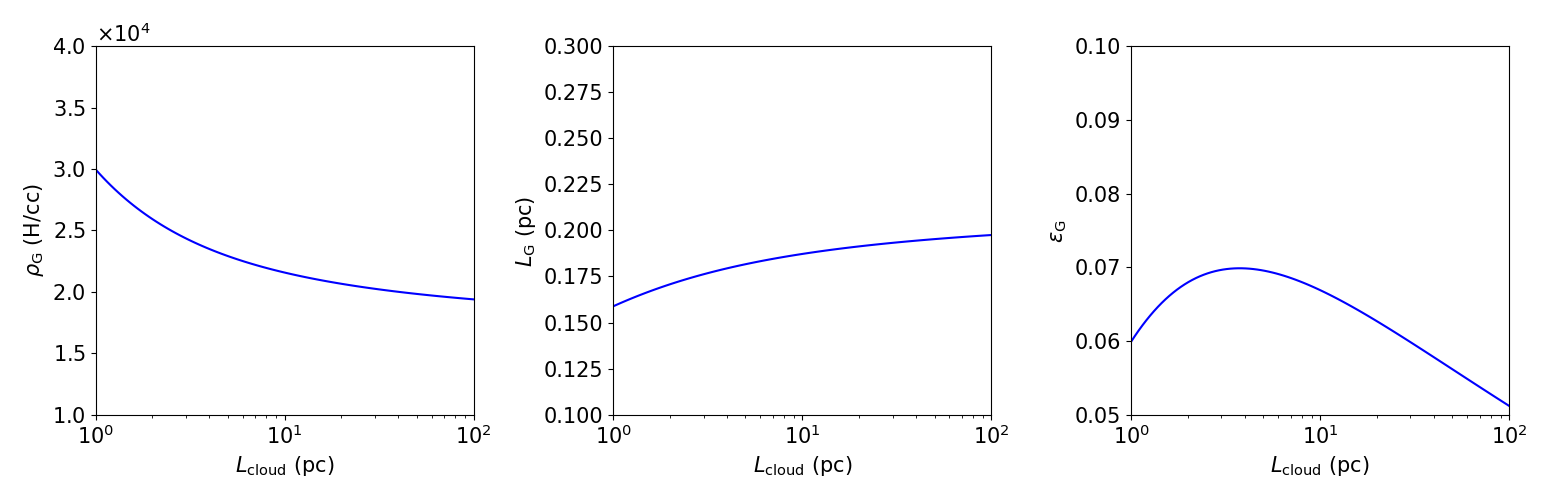}
    \caption{Dependence of the density threshold \(\rho_{\rm G}\) (see Eq. \ref{eq_rho_sg}), the scale threshold \(L_{\rm G}\) (see Eq. \ref{Eq_L_sg}) and the mass fraction of collapsing structures in the molecular gas, \(\epsilon_{\rm G}\), on the size of the parent Larson-type molecular cloud.}
    \label{Fig_L_sg_cloud_dep}
\end{figure*}

For a typical sound speed $c_{\rm s}\approx 0.2\,{\rm km\,s}^{-1}$ ($T=$10 K) this yields a characteristic scale of order $L_{\rm G}\sim 0.2\,\mathrm{pc}$, independently of the cloud size. For typical Larsonian Milky Way-like conditions, the exact dependence of \(\rho_{\rm G}\) and \(L_{\rm G}\) with the size of the molecular clouds is plotted in Fig.~\ref{Fig_L_sg_cloud_dep}. Interestingly, this scale corresponds approximately to the typical width of star-forming filaments  \citep{Arzoumanian_CharacterizingPropertiesNearby2019, Andre_TypicalWidthHerschel2022a, Andre_StructureFragmentationScale2025} and thus to the outermost size of the envelopes of collapsing cores forming within these filaments. 
{Numerous observational studies have reported that the outer radius of the $r^{-2}$ envelope consistently lies close to $0.1\,\rm pc$ \citep{Motte_CircumstellarEnvironmentLowmass2001, Zhang_MassSizeScaling2017, Gieser_PhysicalChemicalComplexity2023, Sahu_ALMASurveyOrion2023, Beuther_DensityDistributionsMagnetic2024}. Notably, \cite{Beuther_DensityDistributionsMagnetic2024} reported that, at scales exceeding $0.1\,\rm pc$, the density profile of the gas surrounding the cores flattens to $r^{-1.5}$, which is consistent with the $k^{-3}$ power spectrum 3D slope assumed in the present model.}

Observationally, this transition is difficult to probe through the density power spectrum because it requires simultaneous access to both the densest regions of molecular clouds and the low-density background over a wide range of spatial scales. In \citet{Schneider_UnderstandingStarFormation2022a}, such a transition is not observed in most of the clouds. We believe that this is due to the fact that the accessible spatial range is too narrow to clearly identify both the \(k^{-3}\) and \(k^{-2}\) slopes. In Appendix \ref{App_PS_grav}, we show that in order to recover the \(k^{-2}\) slope, it is necessary to compute the power spectrum of an \(r^{-2}\) envelope over about four orders of magnitude (\(R_{\rm out}/R\sim 10^4\)). Since the observations of \citet{Schneider_UnderstandingStarFormation2022a} probe the density power spectrum at scales close to the expected transition scale, this explains why the measured slope lies between the two asymptotic values.

{\cite{Andre_StructureFragmentationScale2025} report a transition scale at $0.1\,\rm pc$ in the power spectrum of the column density, shifting from $k^{-1.6}$ at large scales to $k^{-2.8}$ at small scales. According to the slice-projection theorem \citep{Miville-Deschenes_UseFractionalBrownian2003a}, these slopes correspond to those of the 3D power spectrum of the density field, provided that the depth of the structure along the line of sight is at least of the order of the probed modes. Otherwise, they correspond to the 2D power spectrum of the density field. Since the observed structure is thought to be sheet-like, \cite{Andre_StructureFragmentationScale2025} interpret this power spectrum as the 2D power spectrum of the density field. Under this interpretation, the 3D counterparts of these slopes are $k^{-2.6}$ and $k^{-3.8}$, implying a transition in the opposite direction to that assumed in the present model. 
While the line-of-sight depth of the structure is possibly smaller than 1 parsec, it is not expected that it is smaller than the smallest modes probed, i.e. those with a \(10^{-2}\) pc wavelength. Indeed, the cores that are found at small scales are typically isotropic.
We therefore propose that the observed transition reflects a change from sheet-like structures (at scales of $0.1$--$10\,\rm pc$) to three-dimensional structures (at scales of $0.01$--$0.1\,\rm pc$). Under this revised interpretation, the 3D counterparts of the observed slopes would be $k^{-2.6}$ and $k^{-2.8}$, such that the 3D power spectrum exhibits no clear transition. We attribute the absence of an observed transition in the gravity-dominated regime to the fact that the probed spatial range is too narrow to recover the expected $k^{-2}$ slope.}

Numerically, it likewise requires simulations spanning a very broad dynamic range in order to resolve both spectral slopes and to follow the isothermal collapse of dense structures for a sufficiently long time to capture the development of the $\rho\propto r^{-2}$ profiles. 
{In particular, in Appendix~\ref{App_invariant_num_PS}, we show that the power spectra measured in the self-gravitating simulations are not numerically converged because of the limited range of accessible scales. Moreover, the very shallow slope observed at the latest stage of the collapse is likely a consequence of the accumulation of gas at the grid scale. This numerical effect influences not only the smallest scales but also larger scales, extending to the beginning of the inertial range. For this reason, we do not use the slopes of the density power spectra measured in the self-gravitating turbulent simulations to constrain our model.}

{Several recent observational studies have reported infall motions at scales substantially larger than the $L_{\rm G}$ predicted by the present model, typically at the parsec scale \citep{Peretto_GlobalCollapseMolecular2013,Kirk_FilamentaryAccretionFlows2013,  Trevino-Morales_DynamicsClusterformingHubfilament2019, Peretto_StarClusterProgenitors2023}. We emphasise, however, that these observations are not inconsistent with the proposed framework. Indeed, it must be stressed that $\rho_{\rm G}$ does not represent the density threshold above which gravity begins to influence the gas dynamics, but rather the density above which density perturbations are fully shaped by gravity. In particular, $\rho_{\rm G}$ may considerably exceed the mean density of the parent molecular cloud, even if the virial parameter is below one. Similarly, $L_{\rm G}$ should not be interpreted as the largest scale below which gravity begins to affect the flow dynamics, but rather as the scale below which the gas enters a state of near-free fall (despite the pressure support) and develops an $r^{-2}$ density profile. It can therefore be significantly smaller than the scale at which infall motions first appear. The accretion flow predicted by the model from the turbulence-dominated scales to the gravity-dominated cores will be discussed in more detail in Sect.~\ref{Sec_accretion}.}

With this model for the density power spectrum plotted in Fig. \ref{Fig_schema_scenario}, the variance of the density field at a given evolutionary stage of core collapse, corresponding to a characteristic inner radius $R(t)$, can be computed. As shown in Appendix~\ref{App_variance_PS}, one obtains
\begin{multline}
    \label{Eq_Var_PS_tot}
    {\rm Var}\!\left(\frac{\rho}{\bar{\rho}}\right)
    = 32\pi\alpha \\
    \times \left[\frac{1}{3} 
    + \overbrace{\ln\!\left(\frac{L_{\rm inj}}{L_{\rm th}}\right)}^{{\rm WNM }}
    + \overbrace{\ln\!\left(\frac{L_{\rm th}}{L_{\rm cloud}}\right)}^{{\rm CNM }}
    + \overbrace{\ln\!\left(\frac{L_{\rm cloud}}{L_{\rm G}}\right)}^{{\rm turb \ cloud}}
    + \overbrace{\frac{L_{\rm G}}{R(t)} - 1}^{\rm grav \ cloud}\right],
\end{multline}
where the contributions from the neutral gas and molecular gas components of the interstellar medium have been explicitly separated. We recall that \(\bar{\rho}\) is the mean density of kiloparsec scale ISM. We have introduced the parameter \(\alpha=M_{\rm inv}/M_{\rm inj}\) with \(M_{\rm inj}=\bar{\rho}L_{\rm inj}^3\) and \(M_{\rm inv} = {{\rm Var}(\rho)l_{\rm c}^3/\bar{\rho}}\), computed in \cite{Dumond_MassInvariantCompressible2025}.  If one focuses now on the smallest-scale collapsing structures, namely the prestellar cores, of typical size $R$, one can use that $L_{\rm G}/R(t) \gg 1$ and is at least of the same order as $L_{\rm inj}/L_{\rm G}$. Then, the above expression can be simplified to
\begin{equation}
    \label{eq_variance_Rt}
    {\rm Var}\!\left(\frac{\rho}{\bar{\rho}}\right)
    \simeq 32\pi\alpha\,\frac{L_{\rm G}}{R(t)}.
\end{equation}
This result indicates that the variance associated with the gravity-dominated regime dominates the total density variance.
In the following section, we estimate the density variance from the density PDF in order to relate the evolution of the characteristic inner core scale $R(t)$ to the maximum density \(\rho_{\rm max}\) during gravitational collapse.

\subsection{The variance of the density field in the self-gravitating ISM}
\label{Sec_density_Var_PDF}

In order to relate the density variance in a self-gravitating interstellar medium (ISM) to the maximum density of the collapsing cores, one would in principle need a model for the full density PDF, encompassing not only isothermal molecular clouds but the full multiphase ISM. However, such a unified description is currently lacking, despite recent attempts \citep{Colman_RoleTurbulenceSetting2025a}. At present, only the PDF, and its evolution, of isothermal self-gravitating molecular clouds is well described \citep{Jaupart_EvolutionDensityPDF2020}. 

Nevertheless, as shown in the previous section, the density variance is dominated by the gravity-dominated regime, which corresponds to the high-density power-law tail of the PDF. We therefore restrict our analysis to the self-gravitating part of the density PDF, which we model as
\begin{equation}
    \mathcal{P}_\rho(\rho) = B \left(\frac{\rho}{\bar{\rho}}\right)^{-\beta}
    \quad \text{for} \quad \rho \in [\rho_{\rm G}, \rho_{\rm max}],
\end{equation}
where $B$ is a normalisation coefficient that depends on the properties of the multiphase ISM, and $\beta = 2.5$ \citep{Jaupart_EvolutionDensityPDF2020}. As shown in Appendix~\ref{App_variance_PDF} and assuming that all the gravity-dominated gas is contained within clouds, the resulting density variance is then given by
\begin{equation}
    \label{Eq_rho_PDF}
    {\rm Var}\!\left(\frac{\rho}{\bar{\rho}}\right)
    \simeq \epsilon \sqrt{\frac{\rho_{\rm max}\rho_{\rm G}}{\bar{\rho}^2}},
\end{equation}
where $\epsilon=M_{\rm grav}^{\rm cloud}/M_{\rm tot}$ denotes the mass fraction of collapsing gas in a medium of mean density $\bar{\rho}$, i.e. the mass fraction contained in the power law part of the density PDF. 
We decompose this fraction as $\epsilon = \epsilon_{\rm cloud}\epsilon_{\rm G}$, where $\epsilon_{\rm cloud}=M^{\rm cloud}/M_{\rm tot}$ is the molecular gas mass fraction and $\epsilon_{\rm G}=M_{\rm grav}^{\rm cloud}/M^{\rm cloud}$ is the mass fraction of collapsing gas inside molecular clouds. The latter can be computed from the PDF $\mathcal{P}_{\rm cloud}$ (lognormal + power law) derived by \citet{Jaupart_EvolutionDensityPDF2020}, which accurately describes the statistics of self-gravitating isothermal turbulent flows:
\begin{equation}
    \epsilon_{\rm G}
    = \int_{\frac{\rho_{\rm G}}{\bar{\rho}_{\rm cloud}}}^{+\infty}
    \frac{\rho}{\bar{\rho}_{\rm cloud}}
    \mathcal{P}_{\rm cloud}\!\left(\frac{\rho}{\bar{\rho}_{\rm cloud}}\right)
    \, {\rm d}\!\left(\frac{\rho}{\bar{\rho}_{\rm cloud}}\right).
\end{equation}
As shown in Fig. \ref{Fig_L_sg_cloud_dep}, for typical Larsonian molecular clouds, $\epsilon_{\rm G}$ varies only weakly around a value of $\sim 0.06\pm 0.01$. 

Interestingly, as $\epsilon_{\rm G}$ gives an upper bound for the star formation efficiency, the above calculation provides an explanation for the observed inefficiency of the star formation process \citep{Lada_STARFORMATIONRATES2010}.

Then, assuming that the power spectrum of the density field in the turbulent-only ISM is well described by a single power law and that gravity does not affect the amount of molecular gas, we compute in Appendix \ref{App_CNM_mass_fraction} the molecular gas mass fraction. From Eq. \ref{Eq_rho_PDF}, we can then deduce that the variance of the density PDF describing the collapsing gas can be written as:
\begin{equation}
    \label{eq_var_PDF}
    {\rm Var}\!\left(\frac{\rho}{\bar{\rho}}\right)
    \simeq \epsilon_{\rm G} \frac{32\pi\alpha\bar{\rho}\bar{\rho}_{\rm cloud}}{\sigma_{\rm cloud}^2}
    \ln\!\left(\frac{L_{\rm cloud}}{L_{\rm diss}}\right) \sqrt{\frac{\rho_{\rm max}\rho_{\rm G}}{\bar{\rho}^2}},
\end{equation}
where \(\sigma_{\rm cloud}^2\) is the variance of the density field in the molecular clouds and \(L_{\rm diss}\) is the cut-off scale of the density power spectrum in the turbulence-dominated ISM. Note that this scale is likely larger than the core scale reached at the end of the isothermal gravitational collapse, since gravity can form structures significantly smaller than those produced by turbulence alone by efficiently compressing the gas.

\subsection{The characteristic mass of the cores}
\label{Sec_CoreMass}

Using the results of the two previous sections, we can now examine how the characteristic inner radius of the collapsing structures evolves with their central density. Combining Eqs.~\ref{eq_variance_Rt} and \ref{eq_var_PDF}, we obtain
\begin{equation}
    R(t) = \frac{1}{\epsilon_{\rm G}\ln\!\left(\frac{L_{\rm cloud}}{L_{\rm diss}}\right)} 
    \left[b\,\mathcal{M}(L_{\rm cloud})\right]^2
    L_{\rm G}\,
    \frac{\bar{\rho}_{\rm cloud}}{\sqrt{\rho_{\rm max}(t)\rho_{\rm G}}},
\end{equation}
where we have used $\sigma_{\rm cloud}^2/\bar{\rho}_{\rm cloud}^2 = \left[b\,\mathcal{M}(L_{\rm cloud})\right]^2$ (e.g. \citealt{Konstandin_MachNumberStudy2016}).

Using Eq.~\ref{eq_mass_struc_BE}, the characteristic mass of the cores formed in the self-gravitating ISM is thus given by
\begin{multline}
    \label{eq_mean_mass}
    M_{\rm cores} 
    = 4\pi 
    \left[\frac{1}{\epsilon_{\rm G}\ln\!\left(\frac{L_{\rm cloud}}{L_{\rm diss}}\right)}\right]^3 \\ 
    \times \left[b\,\mathcal{M}(L_{\rm cloud})\right]^6
    \left(\frac{\rho_{\rm G}}{\bar{\rho}_{\rm cloud}}\right)^{-2}
    \bar{\rho}_{\rm cloud} \, L_{\rm G}^3 .
\end{multline}

The physical interpretation of this characteristic mass of the cores will be discussed in Sect. \ref{Sec_discussion_mean_mass}.
It is worth at this stage stressing a point of importance. As seen in Eq. \ref{eq_mean_mass}, the maximum density $\rho_{\rm max}(t)$, and thus the time dependence of $ M_{\rm cores}$ disappear in the final result. This is interpreted as the fact that all the mass contained in gravitationally bound structures (i.e. where $\rho>\rho_{\rm G}$), characterised by a collapsing envelope around a central plateau, will end up eventually being contained in the final prestellar cores. This dynamics of the core studied in this model is similar to the one developed by \cite{Larson_NumericalCalculationsDynamics1969} where the density of the central plateau of the core increases while keeping the outer radius \(R_{\rm out}\) fixed.
Such a result implies mass conservation of the collapsing structures, and thus does not take into account outflows or accretion events {that may exist once the collapse has finished, i.e. \(R_{\rm out}\gg R\)}. Although neglecting such dynamical events is certainly a limitation of the present model, it will be shown below and in Sect. \ref{Sec_discussion_mean_mass} that the result of Eq. \ref{eq_mean_mass} is consistent with the mean mass of presently observed CMFs.

Given the number of different parameters entering Eq. \ref{eq_mean_mass}, we may worry that the characteristic mass of the cores in the CMF widely varies with the environment, i.e. the properties of the parent molecular clouds.
In order to test such a behaviour, we consider cloud sizes in the range $1$--$100\,\mathrm{pc}$, corresponding to typical molecular cloud sizes.  
It is important to verify whether this framework predicts a nearly universal characteristic core mass across different (Larsonian) host molecular clouds, since the observed mean core mass typically lies in {a range, $0.5$--$2\,{\rm M}_\odot$, over a wide variety of regions \citep{Zhang_PhysicalPropertiesChemical2018, Konyves_PropertiesDenseCore2020,Takemura_CoreMassFunction2021,Li_ALMASurvey702023a}.} Note that obtaining a robust {observational} estimate of the mean core mass remains challenging because incompleteness at low masses prevents a precise determination of the full core mass function (CMF).

\begin{figure}
    \centering
    \includegraphics[width=\columnwidth]{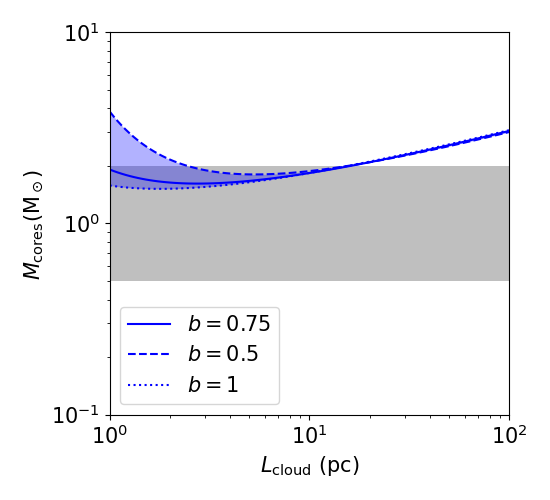}
    \caption{Dependence of the predicted characteristic core mass on the size of the parent Larson-type molecular cloud. The gray region indicates the typical observationally inferred range of mean mass of the cores. The blue region shows the range of predicted characteristic mass of the cores for varying \(b\) between 0.5 (dashed line) and 1 (dotted line).}
    \label{Fig_M_core_cloud_dep}
\end{figure}

Normalising Eq. \ref{eq_mean_mass} to values corresponding to a typical cloud of 20 pc and using the Larson relations, we obtain:
\begin{multline}
    M_{\rm cores} = 2{\rm M}_\odot
    \left(\frac{\epsilon_{\rm G}}{0.07}\right)^{-3}
    \left(\frac{\ln\left(\frac{L_{\rm cloud}}{L_{\rm diss}}\right)}{10}\right)^{-3}
    \left(\frac{b\mathcal{M}_0}{0.75\times 3}\right)^{6} \\
    \times\left(\frac{\rho_{\rm G}}{2\times 10^4{\rm cm}^{-3}}\right)^{-2}
    \left(\frac{d_0}{800 {\rm cm}^{-3}}\right)^{3}
    \left(\frac{L_{\rm G}}{0.2 {\rm pc}}\right)^{3}
    ,
\end{multline}
i.e. \(M_{\rm cores} \in [1.5,\,3]\,{\rm M}_\odot\) for the considered molecular cloud sizes 1 to 100 pc. We introduced \(\mathcal{M}_0=V_0/c_{\rm s}\), the 1~pc normalisation of the Mach number according to the Larson relations. As shown in Fig.~\ref{Fig_M_core_cloud_dep}, this theoretical determination is very consistent with the observational determinations mentioned above. The dependence of this result on the normalisation factors of the Larson relations will be discussed in Sect.~\ref{Sec_Discussion_Star_forming_env}.

Besides the 1~pc normalisation of \(d_0\) and \(V_0\) in the Larson relations (Eq. \ref{eq_larson_velocity} and \ref{eq_larson_density}), Eq. \ref{eq_mean_mass} also involves the dissipation scale $L_{\rm diss}$, but its impact is weak as long as $L_{\rm diss} \ll L_{\rm cloud}$. In this work, we adopt $L_{\rm diss} = 10^{-3}\,\mathrm{pc}$, which is close to the ambipolar diffusion scale in molecular clouds and to the smallest scale at which a power-law density spectrum is still observed \citep{Miville-Deschenes_ProbingInterstellarTurbulence2016a, Pineda_ProbingPhysicsStarFormation2024a}. In the present study, we set the turbulence forcing parameter \(b\) to 0.75 which corresponds approximately to the averaged value observed in star-forming regions \citep{Brunt_DensityVarianceMach2010a, Menon_CompressiveNatureTurbulence2020, Sharda_FirstExtragalacticMeasurement2021,Gerrard_NewMethodSpatially2023, Gerrard_TurbulenceStatisticsClouds2024,  Gerrard_TurbulenceDrivingStarforming2026}. However, according to these studies, \(b\) is observed to vary typically between 0.5 and 1 across star-forming environments.  Considering such a possible variation, the predicted range for the characteristic mass of the cores extends to \([0.2,3]\,{\rm M}_\odot\) as shown in Fig. \ref{Fig_M_core_cloud_dep}, which is still consistent with the typical value reported for Milky Way-like star-forming regions.

Eq.~\ref{eq_mean_mass} shows that the predicted characteristic core mass is independent of the turbulence injection mechanism and of the mean density at the Galactic scale.  Therefore, our theoretical framework provides a solid paradigm to explain the universality of the characteristic core mass in star formation, since the injection scale and mean density may vary significantly across different environments. As dense cores are the direct progenitors of stars, this result naturally provides an explanation for the observed universality of the characteristic stellar mass of the IMF and thus potentially of this latter as a whole.
The dependence of the characteristic core mass on the turbulent forcing parameter $b$ has been investigated numerically by \cite{Mathew_RoleTurbulenceDriving2023} for molecular clouds of about 2~pc, {which reported the same trend as the one predicted by the present model, i.e. that the characteristic mass increases when \(b\) decreases. While the mean mass of the cores they measured in their simulations is a factor of 2 smaller than the one predicted here, there is still a good agreement between this study and the present theory.} 

At this stage, it is important to understand the  nature of the invariant in the general process. As shown initially by Chandrasekhar and generalised by \cite{Jaupart_GeneralizedTransportEquation2021a}, the invariant derives {only} from the equation of mass conservation that couples velocity field and density field, yielding a conservation equation that involves only the integral of the autocovariance function of the {density} field, thus the large-scale normalisation of the density power spectrum, $P_{\rho/\bar{\rho}}({\vec k}=0)$. Since there is no source term in the mass conservation equation, there is indeed no reason for this quantity to vary during the evolution during the star formation process. 
Note that such an invariant does not exist for the velocity field, for which there is no conservation equation without source terms. Therefore, although the value of the invariant in the random density field does not explicitly appear in the final expression for the characteristic core mass, it lies at the very heart of the global physical picture of star formation as it allows us to bridge the gap between turbulence-dominated (\(\propto k^{-3}\)) initial conditions and turbulent (\(\propto k^{-3}\)) and gravity (\(\propto k^{-2}\)) dominated final conditions. 

Physically speaking, this mass invariant corresponds to the characteristic mass of {all} the turbulence-induced correlated structures, which encompass both {over}- and {under}-dense regions. 
However, the mass invariant does not correspond to the mass of the bound cores, as assumed in \cite{Jaupart_GeneralizedTransportEquation2021a}. Going from one to the other requires knowledge of the statistics of the density field affected by gravity, described by \citet{Jaupart_EvolutionDensityPDF2020} (PDF) and Fig.~\ref{Fig_schema_scenario} (density power spectrum). 
In summary, while the invariant is determined solely by the large scales, the mass of the cores is fixed both by this invariant and by the small scales throughout the modification of the statistics of the density field under the influence of gravity.

\section{Discussion}

\subsection{The characteristic mass of the cores across star-forming environments}
\label{Sec_Discussion_Star_forming_env}

\begin{figure}
    \centering
    \includegraphics[width=\columnwidth]{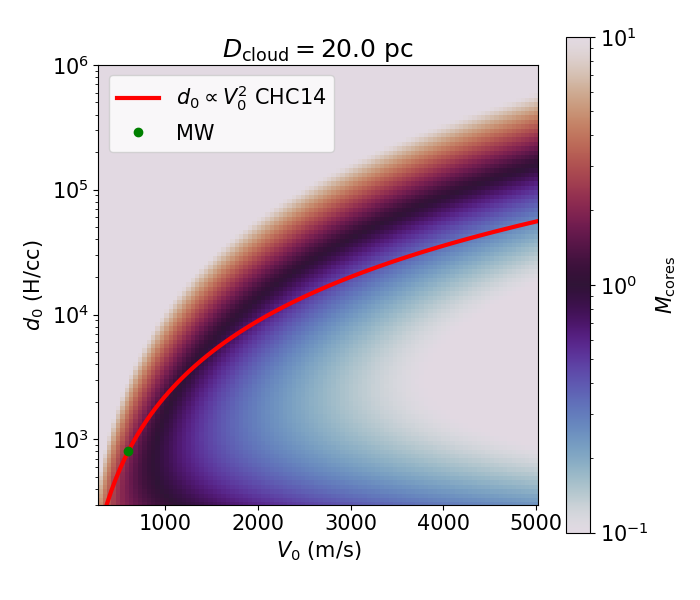}
    \caption{Evolution of the predicted characteristic core mass as a function of the 1-pc normalisation coefficients $d_0$ and $V_0$ of the Larson relations for a molecular cloud of diameter $20\,\mathrm{pc}$. The red curve corresponds to the expected scaling of the Larson relations proposed by \citet{Chabrier_VariationsStellarInitial2014}. The green dot indicates the Larson relation measured in star-forming regions of the solar neighbourhood. For a cloud of size $20\,\mathrm{pc}$ this corresponds to  $M_{\rm cores}\simeq 1\,{\rm M}_\odot$ (see Fig.~\ref{Fig_M_core_cloud_dep}). The uppermost and rightmost values of $d_0$ and $V_0$ along the red line correspond to extreme conditions, notably to massive early-type galaxies, yielding potentially a value $M_{\rm cores}\simeq 0.2\,{\rm M}_\odot$.}
    \label{Fig_M_core_larson_dep}
\end{figure}

The characteristic core mass predicted by Eq.~\ref{eq_mean_mass} depends solely on the normalisation of the Larson relations. The prediction of their variations obtained by the present model is shown in Fig.~\ref{Fig_M_core_larson_dep}. Relative to the fiducial values characteristic of the local interstellar medium, increasing the mean density normalisation $d_0$ (at constant $V_0$) or decreasing the Mach number  normalisation $V_0$ (at constant $d_0$), both of which enhance the relative importance of gravity compared to turbulence, leads to an increase in the predicted characteristic core mass. This trend is consistent with numerous observational studies indicating that high-mass star-forming regions exhibit a CMF with a high-mass slope shallower than the canonical Salpeter value \citep{ONeill_CoreMassFunction2021, Pouteau_ALMAIMFIIIInvestigating2022, Kinman_CoreMassFunction2025}. 

In particular, the ALMA-IMF regions studied by \citet{Koley_ALMAIMFXIXC18O2025} show Mach numbers consistent with those expected from the Larson relations, while their densities are significantly higher. Within our framework, such environments are therefore expected to form systematically more massive cores.

However, except in this specific case, the normalisation coefficients $d_0$ and $V_0$ are generally not independent. As shown by \citet{Chabrier_VariationsStellarInitial2014}, both are determined by the mean external pressure of the medium, assuming that turbulence is primarily driven by large-scale Toomre instability. Under this assumption, they find the relation
\begin{equation}
    \frac{d_0}{d_{\rm MW}} = \left(\frac{V_0}{V_{\rm MW}}\right)^2\propto \left(\frac{P^{\rm ext}}{P^{\rm ext}_{\rm MW}}\right)^{1/2} ,
\end{equation}
where $d_{\rm MW}$ and $V_{\rm MW}$ are the standard normalisations in usual MW-like environments, and \(P^{\rm ext}\) is the external pressure. This implies that typical star-forming regions should lie along the red curve shown in Fig.~\ref{Fig_M_core_larson_dep}. Along this sequence, increasing $d_0$ and $V_0$ consistently leads to a decrease in the predicted characteristic core mass relative to the one observed in Milky Way-like star-forming regions.
This is  expected to correspond to massive early-type galaxies (ETGs), which were significantly denser and more turbulent than present-day spiral galaxies \citep{Genzel_RingsBulgesEvidence2008, VanDokkum_ConfirmationRemarkableCompactness2008}. Although it is currently impossible to directly measure the core mass distribution in such systems, recent IMF studies suggest an excess of low-mass stars compared to the Milky Way IMF \citep[see][for a review]{Smith_EvidenceInitialMass2020a}. This indeed implies a lower mean stellar mass in these galaxies. 
Our model predictions, yielding a smaller mean core mass in massive ETGs than in the MW, are in qualitative agreement with these observations, provided that there is a direct correspondence between the dense cores and the emerging  stars.

\subsection{The characteristic mass of the cores and the mean mass of the CMF}
\label{Sec_discussion_mean_mass}

\begin{figure}
    \centering
    \includegraphics[width=\columnwidth]{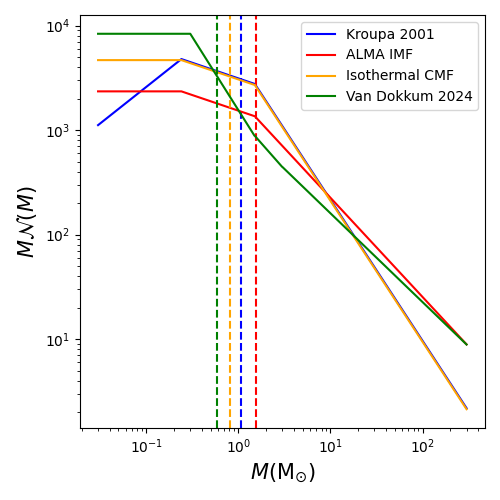}
    \caption{Comparison between mean masses computed from different core mass functions (CMFs).}
    \label{fig_CMF_mean_mass}
\end{figure}

Fundamentally, the quantity given by Eq.~\ref{eq_mean_mass} corresponds to the characteristic mass of bound structures formed in a self-gravitating turbulent medium.
Its derivation is based on a model of typical collapsing structures, characterised by an inner radius $R(t)$ that decreases with time and an outer density threshold $\rho_{\rm G}$. This density corresponds to the characteristic value above which the gas becomes gravitationally unstable and collapses. 

Before comparing this value with the mean mass in various mass distributions, it is important to stress that this mass should not be interpreted as either the peak or the median mass of the CMF.
Indeed, the collapse considered here is isothermal: we follow the evolution of the density fluctuations prior to the transition to the adiabatic regime, i.e.  before the gas reaches the adiabatic density threshold. As a result, the collapse is self-similar and no physical process exists to halt fragmentation, which implies a power-law distribution across the mass spectrum. Consequently, no characteristic peak is expected to emerge in the mass distribution of cores formed during an isothermal collapse. For collapsing structures, i.e. $\rho(r)\propto r^{-2}$, the CMF is expected to be flat at low masses, i.e. $M\,\mathcal{N}(M) \propto M^0$, where $\mathcal{N}(M)=dN/dM$ denotes the number of cores per mass interval \citep{Lee_StellarMassSpectrum2018}.

In Fig.~\ref{fig_CMF_mean_mass}, we plot the { mean mass}  computed for different CMF models. 
All these CMFs have a functional form typical of the fiducial one in star formation, i.e. a Salpeter-like high-mass slope rolling down to a flattish or decreasing form below some mass. For the sake of simplicity, we adopt a Kroupa-like multi-power-laws parametrisation \citep{Kroupa_VariationInitialMass2001}, shifted by a core-to-star efficiency factor of 3 (see \citealt{Houghton_CoresStarsSearching2024} and references therein).
Although such a parametrisation has no physical justification, we use it rather than the \citet{Chabrier_GalacticStellarSubstellar2003} form for illustrative purposes because it is more straightforward to modify it in order to introduce a flat power law at low masses.

The CMF expected for an isothermal collapse is shown in orange. 
As illustrated in Fig. \ref{fig_CMF_mean_mass}, we find that replacing the usual decreasing low-mass distribution by a flat one modifies the mean mass by only $\sim20\%$. In all cases, the mean mass of the CMF remains similar to the value derived from Eq.~\ref{eq_mean_mass} for the characteristic mass. The same result holds for the mean mass value derived from the Chabrier 2005 IMF shifted by a factor of 3 (i.e. $\simeq 1.6\,{\rm M}_\odot$). Therefore, {for such standard CMF-type functional forms}, there is a close correspondence between the mean mass of the cores and the characteristic mass computed from Eq.~\ref{eq_mean_mass}.

We also compare with CMFs representative of high-mass star-forming regions and massive ETGs. The former is constructed by adopting the shallower high-mass slope observed in the ALMA-IMF survey \citep{Pouteau_ALMAIMFIIIInvestigating2022, Louvet_ALMAIMFXVCore2024}. The latter corresponds to the CMF inferred for early-type galaxies by \citet{VanDokkum_ReconcilingRatiosCosmic2024}. As discussed in Sect.~\ref{Sec_Discussion_Star_forming_env} and predicted by our model, the mean core mass is expected to be smaller in early-type galaxies than in Milky Way-like star-forming regions, while it is expected to be larger in massive protoclusters such as those probed by the ALMA-IMF survey.

\subsection{The Larson relations}

Since the Larson relations are one of the key ingredients of the present model, it is important to recall our current physical understanding of them. 
First, as shown in \cite{Hennebelle_AnalyticalTheoryInitial2008}, the scaling of the velocity--size relation (\(\sigma_{\rm 3D}\propto R^{\eta_v}\), see Eq. \ref{eq_larson_velocity}) arises directly from the power spectrum of the velocity field in supersonic compressible turbulent flows. The 3D index \(n\) of the velocity power spectrum can be related to \(\eta_v\) through \(\eta_v=(n-3)/2\). For \(n=4\), expected for shock-dominated turbulence \citep{Burgers_MathematicalModelIllustrating1948} and confirmed numerically \citep{Federrath_ComparingStatisticsInterstellar2010,Federrath_UniversalitySupersonicTurbulence2013, Fensch_UniversalGravitydrivenIsothermal2023}, we obtain \(\eta_v=0.5\). The value of the one-parsec normalisation \(V_0\) has been suggested to be set by vertical hydrostatic equilibrium at the scale of the galactic disk, leading to a 3D normalisation of \(600\,{\rm m\,s^{-1}}\) \citep{Chabrier_VariationsStellarInitial2014}\footnote{The normalisation \(V_0\) is defined here for a cloud characterised by a one-parsec diameter, while in \citet{Chabrier_VariationsStellarInitial2014} it is defined for a one-parsec radius. The two values therefore differ by a factor \(\sqrt{2}\).}. This value is consistent with recent observations of molecular clouds \citep{Miville-Deschenes_PHYSICALPROPERTIESMOLECULAR2017, Benedettini_MolecularCloudCatalogue2021, Feng_RevisitingVelocityDispersion2024}.

The scaling of the density--size relation is understood as a consequence of the virialisation of molecular clouds \citep{McKee_TheoryStarFormation2007}. The virial parameter is defined as:
\begin{equation}
    \alpha_{\rm vir}=\frac{2 E_{\rm kin}}{E_{\rm grav}}=\frac{5\sigma_{3D}^2}{\pi G L_{\rm cloud}^2 \bar{\rho}_{\rm cloud}},
\end{equation}
for a spherical homogeneous cloud of diameter \(L_{\rm cloud}\) and 3D velocity dispersion \(\sigma_{\rm 3D}\). Using the velocity--size relation, we obtain:
\begin{equation}
    d_0=\frac{5V_0^2}{\pi G (1{\rm pc})^2\alpha_{\rm vir}}.
\end{equation}
For \(\alpha_{\rm vir}\sim 3\), which corresponds to the typical value of the virial parameter measured in molecular clouds \citep{Miville-Deschenes_PHYSICALPROPERTIESMOLECULAR2017, Colombo_IntegratedPropertiesMolecular2019, Benedettini_MolecularCloudCatalogue2021, Feng_RevisitingVelocityDispersion2024}, we obtain \(d_0=800\,{\rm cm}^{-3}\) which is the fiducial value of the present model, corresponding to a typical surface density of \(\Sigma_0 = 30\,{\rm M}_\odot\,{\rm pc}^{-2}\).

There is a significant scatter in the measured virial parameter and mean surface density of molecular clouds, which results in uncertainties in the determination of \(V_0\) and \(\Sigma_0\). However, as shown in Fig.~\ref{Fig_M_core_larson_dep}, varying these normalisations by a factor of a few around the fiducial values used in the present study does not lead to a significant change in the predicted characteristic mass of the cores.

{
\subsection{Accretion rate of prestellar cores}
\label{Sec_accretion}

The bulk of the mass constituting the final prestellar cores originates from regions in which the initial density is below $\rho_{\rm G}$. This gives rise to an accretion flow from the turbulent medium, consistent with the infalling gas observed around prestellar cores along filaments on scales of approximately $0.1$--$1\,\rm pc$ \citep{Peretto_GlobalCollapseMolecular2013, Kirk_FilamentaryAccretionFlows2013, Trevino-Morales_DynamicsClusterformingHubfilament2019, Peretto_StarClusterProgenitors2023}. The mean accretion rate driven by gravity is estimated as the difference between the mass accumulated in gas denser than $\rho_{\rm G}$ once gravity has formed dense cores with $\rho_{\rm max} \gg \rho_{\rm G}$ (described by a power-law PDF) and the mass present at densities above $\rho_{\rm G}$ in the purely turbulent case (described by a lognormal PDF), divided by the number of cores formed $N_{\rm cores}$ and the free-fall timescale evaluated at $\rho_{\rm G}$. This yields:
\begin{equation}
    \dot{M} = L_{\rm cloud}^3\bar{\rho}_{\rm cloud}\frac{\int_{\rho_{\rm G}}^{+\infty}\rho\,\mathcal{P}_{\rm cloud}(\rho)\,{\rm d}\rho-\int_{\rho_{\rm G}}^{+\infty}\rho\,\mathcal{P}_{\rm turb}(\rho)\,{\rm d}\rho}{N_{\rm cores}\, \tau_{\rm ff}(\rho_{\rm G})},
\end{equation}
where $\tau_{\rm ff}(\rho_{\rm G})=\sqrt{3\pi/(32\rho_{\rm G}G)}$ is the free-fall timescale at density $\rho_{\rm G}$. For simplicity, we assume that all cores have the same mass, which implies $N_{\rm cores}=\epsilon_{\rm G} L_{\rm cloud}^3\bar{\rho}_{\rm cloud}/M_{\rm core}$. Here, $\mathcal{P}_{\rm cloud}$ denotes the lognormal plus power-law PDF introduced by \cite{Jaupart_EvolutionDensityPDF2020}, and $\mathcal{P}_{\rm turb}$ is the lognormal PDF describing the gas prior to any gravitational modification of its density statistics. For Milky Way-like conditions, $\dot{M}$ is found to range typically between $5\times 10^{-6}\,{\rm M}_\odot\,{\rm yr}^{-1}$ and $10^{-5}\,{\rm M}_\odot\,{\rm yr}^{-1}$, in agreement with accretion rates measured in the vicinity of solar-mass cores \citep{Lee_SurveyInfallMotions2001, Kim_RoleFilamentaryStructures2022, Moon_PrestellarCoresTurbulent2025, Morii_ALMASurvey702026}. This result demonstrates that the present model is fully consistent with gravity-driven infall at densities below $\rho_{\rm G}$ (and on scales larger than $L_{\rm G}$), as commonly observed in star-forming regions.}

\subsection{Limitations of the present study}
\label{Sec_limitation}

Our model relies on several key assumptions that merit discussion. First, the existence of a temporal invariant at galactic scales requires that the medium be {statistically} homogeneous. While this condition is clearly not satisfied if the system under consideration is the entire galaxy, it is likely to be reasonably valid for a sub-volume of size $\mathcal{V} = 1\,\mathrm{kpc} \times 1\,\mathrm{kpc} \times 100\,\mathrm{pc}$, where the $100\,\mathrm{pc}$ dimension corresponds to the direction perpendicular to the Galactic plane. 
Indeed, the characteristic size required to ensure a sufficient degree of statistical homogeneity should be comparable with the width of a spiral arm in the Galactic plane and smaller than the Galactic scale height in the vertical direction.  
Such a volume is expected to be subject to significant shear. However, as shown by \citet{Jaupart_GeneralizedTransportEquation2021a}, the temporal invariance of $M_{\rm inv}$ is preserved in linear shear flows and thus is not violated by galactic differential rotation.

The existence of the invariant further requires the presence of an injection scale, i.e. a scale above which the gas statistics become uncorrelated. Although the final results do not depend explicitly on the value of this scale, the model nevertheless assumes its existence. In particular, this implies that the volume $\mathcal{V}$ should not be undergoing strong coherent accretion, which is a good assumption in present-day spiral galaxies \citep{Sancisi_ColdGasAccretion2008}. In contrast, gas accretion from the intergalactic medium onto galactic discs is expected to be significantly stronger in high-redshift galaxies \citep{Dekel_ColdStreamsEarly2009}, making the applicability of the present framework to such systems more uncertain.

Another important assumption concerns the functional form of the density power spectrum. First, we assume the existence of a single turbulent cascade extending continuously across the neutral gas and the molecular gas, despite the sharp density transition at the cooling scale produced by thermal instability. While we do not currently have a first-principles physical explanation for this behaviour, it is strongly supported by a wide range of observational studies \citep{Martin_GHIGLSMappingIntermediate2015, Marchal_ROHSARegularizedOptimization2019, Marchal_ThermalTurbulentProperties2021, Elmegreen_PowerSpectraJWST2025}. Secondly, the predicted characteristic mass depends on the assumed slope of the density power spectrum in the turbulence-dominated regime. Based on the model proposed by \cite{Dumond_SlopePowerSpectrum2026}, together with the density variances inferred in the ISM \citep{Gerrard_TurbulenceDrivingStarforming2026} and the fact that the inertial range extends over at least four orders of magnitude, the density power spectrum is expected to exhibit slope exponents between $-2.9$ and $-3.1$. Such variations around the fiducial $k^{-3}$ scaling adopted in the present work modify the predicted characteristic masses by at most a factor of $\sim 3$. Given the uncertainties in the mass distribution of the cores, the prediction of the model still remains consistent with the current constraints on the characteristic mass of the cores.

{Finally, we have assumed that Larson's relations hold from $100\,\rm pc$ down to $0.1\,\rm pc$. However, since infall motions are observed on scales typically ranging from $1\,\rm pc$ to $10\,\rm pc$ \citep{Peretto_StarClusterProgenitors2023}, it is possible that such motions amplify the turbulent velocity field beyond the levels expected from Larson's relations \citep{Robertson_ADIABATICHEATINGCONTRACTING2012}. Indeed, in a collapsing medium, the turbulent velocity scaling relation is expected to saturate or even reverse in sign. This could in turn lead to an increase in the value of $\rho_{\rm G}$. However, given that subsonic and transonic structures are observed within cores at the scales predicted by Larson's relations, this suggests that such a mechanism does not have sufficient time to operate during the early stages of star formation.}

\section{Conclusion}

In this paper, we have investigated the role of the Chandrasekhar invariant, reformulated by \citet{Jaupart_GeneralizedTransportEquation2021a}, in the astrophysical context of star formation. Using high-resolution numerical simulations, we first demonstrated that this invariant is indeed preserved in an isothermal self-gravitating medium. Based on the assumptions underlying its derivation, we further argue that the thermodynamic properties of the gas should not violate this invariance. As a result, this quantity is expected to be conserved throughout the entire star formation process. 
Using the alternative formulation of the invariant, namely the temporal invariance of the zero-wavenumber value of the density power spectrum, we showed that it can be used to predict the characteristic mass of star-forming cores. This derivation relies on a model for the density power spectrum, which we constrain using both theoretical considerations and observational evidence. We find that the resulting characteristic mass depends only on the normalisation of the Larson relations. 
This provides a strong argument in favour of the universality of the mean core mass, and consequently of the mean stellar mass. For Milky Way-like star-forming regions, the predicted value is in very good agreement with the observed mean core mass.
By examining how the normalisation of the Larson relations may vary across different star-forming environments, we further show that this framework naturally accounts for the observed variations in the characteristic core mass in high-mass star-forming regions and in early-type galaxies.
Altogether, this picture suggests that the characteristic core mass is determined both by the initial conditions of the star formation process, namely the turbulent, multiphase interstellar medium, and by the statistics of the small-scale collapsing structures.

In summary, the existence of such a mass time-invariant during the ongoing star formation process, a direct consequence of the universal statistical  properties of the random field of density fluctuations generated by turbulence across all scales from galactic ones (the injection scale) to prestellar core ones,  represents a fundamental quantity in our quest to understand star formation and the origin of the IMF.
Finally, going back to the heated debate between ``turbulence'' and ``gravity''-dominated star formation scenarios mentioned in Sect. \ref{Introduction}, the present paradigm shows that both (universal) processes play a joint role during the evolution of gravo-turbulent molecular clouds in shaping the final CMF/IMF, although at different times and scales.

\begin{acknowledgements} 
   We thank the referee Christoph Federrath for his constructive report that helped to improve the clarity of this work. 
   We thank No\'e Brucy, Dong Lai, Patrick Hennebelle and Guillaume Laibe for very helpful discussions that improved the clarity of the manuscript. Simulations were produced using GENCI allocations (grants A0150411111 and A0170411111). We gratefully acknowledge support from the CBPsmn (PSMN, Pôle Scientifique de Modélisation Numérique) of the ENS de Lyon for the computing resources. The platform operates the SIDUS solution \citep{Quemener_SIDUS} developed by Emmanuel Quemener. PD and JF were supported by the French national research agency grant ANR-23-CE31-0005 (BRIDGES).
\end{acknowledgements}

\bibliographystyle{aa}
\bibliography{Bib}

\begin{appendix}

    \section{Numerical convergence}
    \label{App_convergence}
\begin{figure*}[h!]
        \centering
        \sidecaption
        \includegraphics[width=12cm]{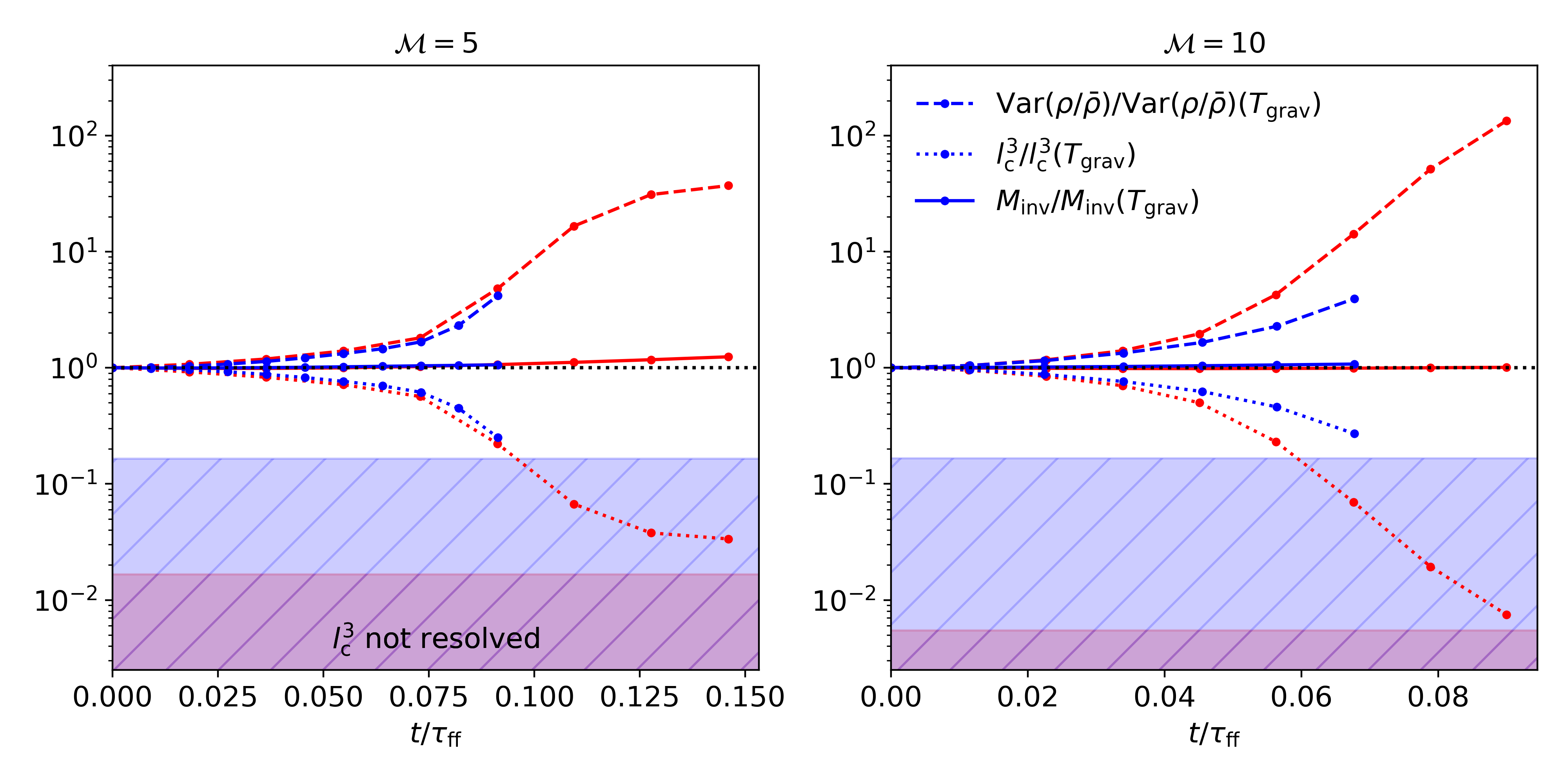}
        \caption{Same as Fig. \ref{fig_M_inv_grav} for two different resolutions, \(2048^3\) in red and \(1024^3\) in blue. The red and blue shaded areas correspond to the unresolved values of the correlation length for the \(2048^3\) and \(1024^3\) resolutions respectively.}
        \label{fig_M_inv_grav_conv}
     \end{figure*}

    In this appendix, we study how the time independence of the invariant is affected by numerical resolution. In Fig. \ref{fig_M_inv_grav_conv}, we plot the evolution of the invariant for two different resolutions, \(1024^3\) (blue) and \(2048^3\) (red). 
    The evolutions of the variance and of the correlation length are not converged at Mach 10. As already shown in \cite{Dumond_MassInvariantCompressible2025}, increasing the Mach number makes the conservation of the invariant and the correlation length harder. However, we see that the time invariance of the \(M_{\rm inv}\) is not affected by the resolution. Increasing the resolution allows one to follow the invariant during a longer time. 
    
     \section{Invariance of the zero-wavenumber value of the density power spectrum}
     \label{App_invariant_num_PS}
     {
     In this appendix, we study an alternative formulation of the invariant, namely that the zero-wavenumber value of the density power spectrum remains constant in time if the mean density of the medium is also constant. 
     Unfortunately, the usual way to estimate the power spectrum using a Fourier transform of the density field is a biased and unreliable estimator at zero wave number, since it enforces its value to be zero by construction. One can however look at the largest scales that are weakly affected by this bias, typically those around and just above \(L_{\rm inj}\), to infer the value of the power spectrum at zero wavenumber. 
     In Fig.~\ref{fig_M_inv_PS_conv}, we show that the value of the power spectrum at these largest accessible scales remains constant during the collapse. Moreover, while this value is also independent of numerical resolution, the power spectrum is not converged at scales smaller than the injection scale. The inertial range is systematically shallower at lower resolution, following the same trend as that predicted by \citet{Dumond_SlopePowerSpectrum2026}. This suggests that the slope of the power spectrum measured in numerical simulations at any scale smaller than the injection scale depends on the numerical resolution. For this reason, we do not rely on the slopes of the density power spectra measured in the simulations to construct our model.
     
     \begin{figure*}
        \sidecaption
         \centering
         \includegraphics[width=12cm]{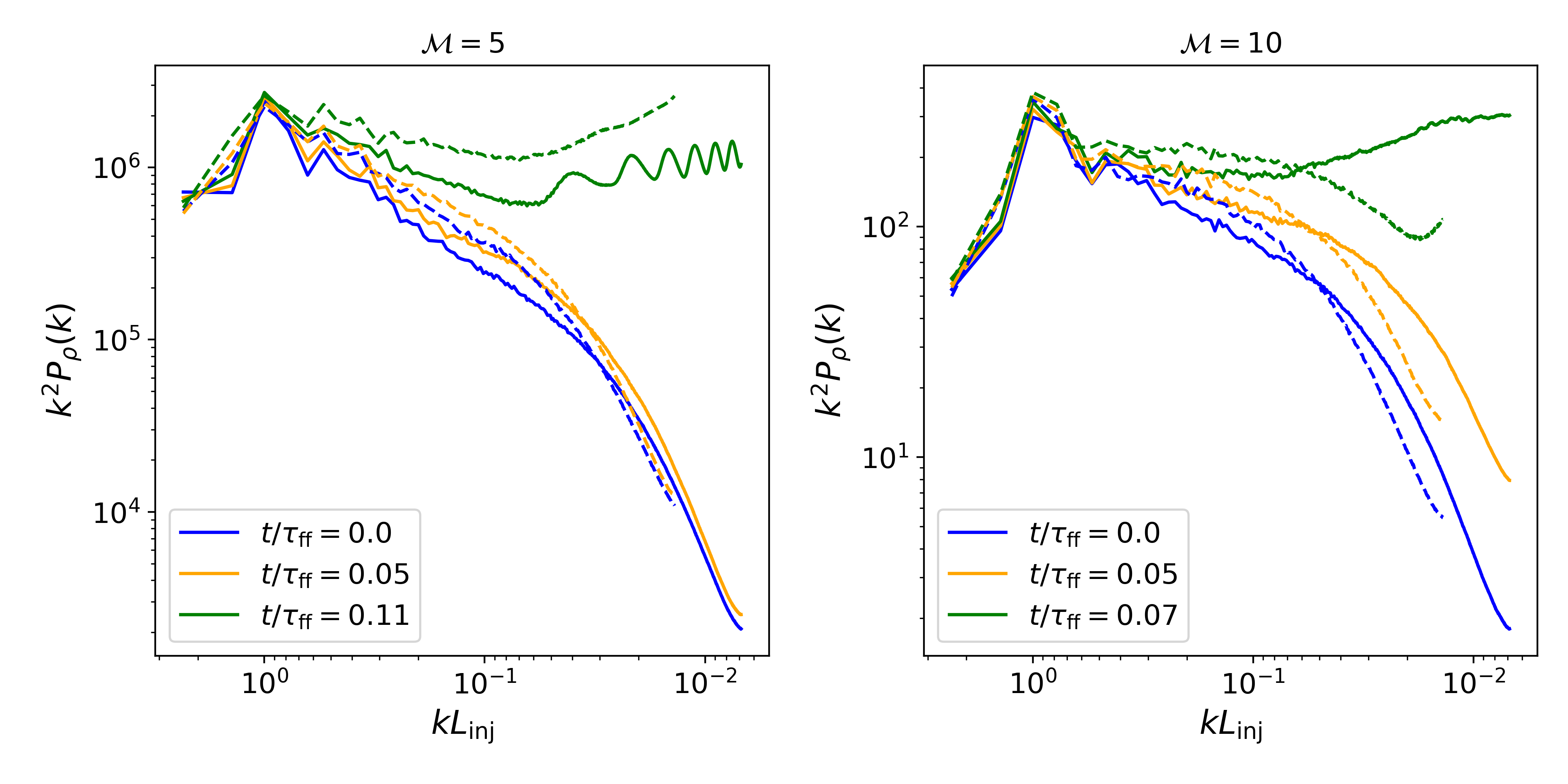}
         \caption{Power spectrum of the density field at different stages of the collapse at resolutions of \(2048^3\) (solid lines) and \(1024^3\) (dashed lines). While the inertial-range spectrum evolves significantly during the collapse, the value of the power spectrum at the smallest accessible mode remains constant with both time and numerical resolution.}
         \label{fig_M_inv_PS_conv}
     \end{figure*}
     }
    \section{The power spectrum of self-gravitating turbulence}
    \label{App_PS_grav}
    
    In this appendix, we demonstrate the link between the density profile of collapsing structures and the slope of the density power spectrum at the scales at which these structures exist. 
    We consider the following functional form for the density power spectrum:
    \begin{equation}
        \label{Eq_functional_form_PS}
        P_{\frac{\rho}{\bar{\rho}}}(k) = \frac{A}{\left(1+\left(\frac{k L_{\rm inj}}{2\pi}\right)^2\right)^{\eta/2}}{\rm H}(1-\frac{k R}{2\pi}), 
    \end{equation}
    where \(H\) is the Heaviside function.
    We especially assume a sharp cut-off at the scale \(R\) which corresponds to the inner scale of the dense cores. This assumes that the smallest structures formed in the ISM by gravity are the cores. Smaller scale structures than the cores may form by turbulence but they will not be gravitationally bound, and they will be characterised typically by a \(k^{-3}\) slope. The contribution of this regime to the total variance will be negligible compared to the gravity dominated domain. We will thus neglect any contribution in the power spectrum by any scales smaller than the typical inner scale of the cores. 
    
    In the following, we predict the power spectrum of the density field made of collapsing density structures unaffected by further (negligible) gas accretion once they form and characterised by a radial profile given by \(\rho(r)\propto r^{-\kappa}\). As the collapse proceeds, i.e. before reaching the density of the first Larson core, the internal radius \(R\) decreases and the density \(\rho_{\rm max}\) increases such that
    \begin{equation}
        R^3\propto \rho_{\rm max}^{-3/\kappa}.
    \end{equation}
    The non-accreting condition on which this relation is based is justified by the fact that the collapse timescale of the inner part of the structure is much shorter than the collapse timescale of the outer part of the structure. 
    
    Moreover, given the functional form of the density power spectrum given by Eq. \ref{Eq_functional_form_PS}, we have:
    \begin{equation}
        {\rm Var}(\rho)\propto \int_{0}^{L_{\rm inj}/R}\frac{u^2}{\left(1+u^2\right)^{\eta/2}}{\rm d}u.
    \end{equation}
    In the limit \(R\ll L_{\rm inj}\), we have:
    \begin{equation}
        {\rm Var}(\rho)\propto R^{\eta - 3},
    \end{equation}
    if  \(\eta\neq 1\). Finally, from the density profile of the collapsing structure, we have:
    \begin{equation}
        {\rm Var}(\rho)\propto \rho_{\rm max}^{\frac{3-\eta}{\kappa}}.
    \end{equation}
    
\begin{figure*}
        \centering
        \sidecaption
        \includegraphics[width=6cm]{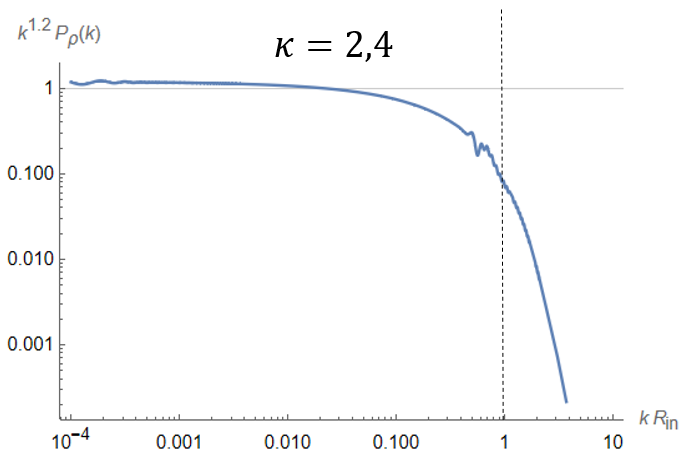}
        \includegraphics[width=6cm]{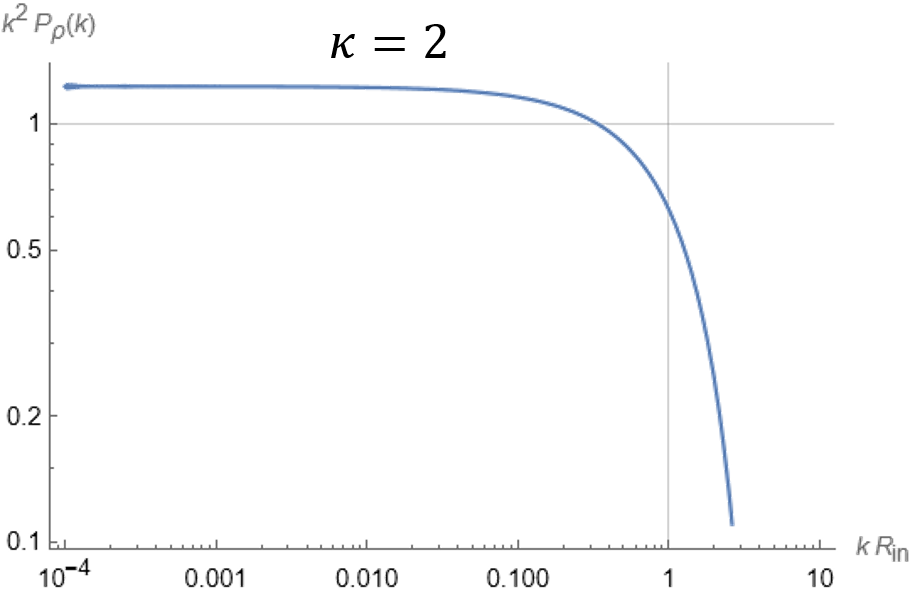}
        \caption{Numerical verification of Eq. \ref{eq_powerspectrum_cores} for two density profiles, \(\kappa=2.4\) (left) and \(\kappa=2\) (right). The power spectra are computed as the square of the Fourier transform of the density profile \(|\hat{\rho}(k)|^2\). The density envelope characterised by a slope \(\kappa\) starts at the inner radius \(R_{\rm in}\).}
        \label{Fig_power_spectrum_cores}
    \end{figure*}

    Moreover, calculating the density variance with a PDF with a high-density slope \(\mathcal{P}\propto\rho^{-(1+\frac{3}{2})}\), corresponding to density with core having a radial profile \(r^{-\kappa}\) \citep{Federrath_STARFORMATIONEFFICIENCY2013}, the slope of the density profile can be linked to the variance by:
    \begin{equation}
        {\rm Var}(\rho) \propto \rho_{\rm max}^{2-\frac{3}{\kappa}}.
    \end{equation}
    Thus, we get
    \begin{align}
        \label{eq_powerspectrum_cores}
        \eta &= 6-2\kappa,
    \end{align}
    for \(\kappa\in]1.5, 2.5[\).
    For \(\kappa=2\), we get \(\eta=2\). This means that, given the density profile of collapsing structures, the power spectrum in a collapsing medium should flatten to \(\eta=2\), corresponding to a -2 slope in 3D. This relation is verified numerically in Fig. \ref{Fig_power_spectrum_cores} for two density profiles, \(\kappa=2\) and \(\kappa=2.4\).

    \section{Universality of \(L_{\rm G}\) for Larsonian star-forming molecular clouds}
    \label{App_unviserval_L_grav}
    
    In this appendix, we explain why the critical density \(\rho_{\rm G}\) given by Eq. \ref{eq_rho_sg} following \cite{Jaupart_EvolutionDensityPDF2020}, and thus the critical scale \(L_{\rm G}\) below which gravity dominates, does not depend on the size of the parent Larsonian molecular cloud. 
    As shown in \cite{Jaupart_EvolutionDensityPDF2020}, Eq. \ref{eq_rho_sg} can be rewritten on the following form:
    \begin{equation}
        \frac{\rho_{\rm G}}{\bar{\rho}_{\rm cloud}}-1 =b^2\mathcal{M}^2\alpha_{\rm vir}\frac{\ln(\rho_{\rm G}/\bar{\rho}_{\rm cloud})+\sigma_s^2/2}{\sigma_s^2},
    \end{equation}
    where \(\alpha_{\rm vir}\) is the virial parameter of the Larsonian molecular clouds. The Larson relations impose that the virial parameter is constant and close to 1 as often observed (see e.g. \cite{Peretto_StarClusterProgenitors2023}). 
    Thus, as \(\rho_{\rm G}\) is large compared to the mean density of the cloud and using once again the Larson relation, it follows:
    \begin{equation}
        \rho_{\rm G} = d_0b^2\mathcal{M}_0^2\left(\frac{\ln(\rho_{\rm G}/d_0) + \ln(L_{\rm 1pc})}{\ln((b\mathcal{M}_0)^2)+\ln(L_{\rm 1pc})}+\frac{1}{2}\right),
        \label{rho}
    \end{equation}
    where we introduce \(d_0\) and \(\mathcal{M}_0\) respectively the density and the Mach number at the scale of 1~pc as given by the normalisations in Eqs.~\ref{eq_larson_velocity} and \ref{eq_larson_density} and \(L_{\rm 1pc}\) the size of the cloud normalised at 1~pc. A possible, approximate  solution for Eq. \ref{rho} is \(\rho_{\rm G} \simeq d_0b^2\mathcal{M}_0^2\). It is most likely a unique solution for $ {\rho_{\rm G}}/{\bar{\rho}_{\rm cloud}}>1$, which is the domain we are interested in.
    Note that this solution is exact if the factor \(1/2\) were negligible compared to the first term, which is not strictly the case. 
    Strictly speaking, \(\rho_{\rm G}\) still depends very weakly on the size of the parent molecular cloud, making \(L_{\rm G}\) vary by less than 10\% for molecular clouds with sizes ranging from 1~pc to 100~pc. We have properly taken this into account in Fig. \ref{Fig_M_core_cloud_dep}.

    \section{The density variance from the power spectrum}
    \label{App_variance_PS}
    Based on the arguments given in Sect.~\ref{Sec_density_PS}, our model for the density power spectrum describing the two-point statistics of the interstellar medium is therefore given by
    \begin{equation}
        \label{Eq_functional_form_PS_tot}
        P_{\frac{\rho}{\bar{\rho}}}(k) = {\rm Var}\!\left(\frac{\rho}{\bar{\rho}}\right)
        \begin{cases}
            A, & \dfrac{kL_{\rm inj}}{2\pi} < 1, \\
            A \left(\dfrac{kL_{\rm inj}}{2\pi}\right)^{-3}, & 1 < \dfrac{kL_{\rm inj}}{2\pi} < \dfrac{L_{\rm inj}}{L_{\rm G}}, \\
            A \left(\dfrac{L_{\rm inj}}{L_{\rm G}}\right)^{-3}
            \left(\dfrac{kL_{\rm G}}{2\pi}\right)^{-2},
            & 1 < \dfrac{kL_{\rm G}}{2\pi} < \dfrac{L_{\rm G}}{R(t)}.
        \end{cases}
    \end{equation}
    
    The variance of the density field described by this power spectrum is thus given by:
    \begin{align}
        &{\rm Var}\left(\frac{\rho}{\bar{\rho}}\right) =\frac{4\pi}{(2\pi)^3} \int_{0}^{+\infty}P_{\frac{\rho}{\bar{\rho}}}(k)k^2{\rm d}k \\
        &= 32\pi{\rm Var}\left(\frac{\rho}{\bar{\rho}}\right)\frac{ l_{\rm c}^3}{L_{\rm inj}^3}\left(\int_{0}^{1}u^2{\rm d}u + \int_{1}^{\frac{L_{\rm inj}}{L_{\rm G}}}\frac{{\rm d}u}{u}+\int_{\frac{L_{\rm inj}}{L_{\rm G}}}^{\frac{L_{\rm inj}}{R}}\left(\frac{L_{\rm inj}}{L_{\rm G}}\right)^{-1}{\rm d}u\right).
    \end{align}
    We introduce \(\alpha=M_{\rm inv}/M_{\rm inj}\) with \(M_{\rm inj}=\bar{\rho}L_{\rm inj}^3\) and \(M_{\rm inv} = {\rm Var(\rho)l_{\rm c}^3/\bar{\rho}}\) already computed in \citet{Dumond_MassInvariantCompressible2025}. We thus find the expression given in Eq. \ref{Eq_Var_PS_tot}:
    \begin{equation}
        {\rm Var}\left(\frac{\rho}{\bar{\rho}}\right) = 32\pi\alpha\left[1/3 + \ln(\frac{L_{\rm inj}}{L_{\rm G}})+\frac{L_{\rm G}}{R}-1\right].
    \end{equation}

    \section{The density variance from a power-law-like PDF}
    \label{App_variance_PDF}
    In this appendix, we compute the variance of the density field from the density PDF. We assume that the PDF can be modelled by a power law above a given density threshold \(\rho_{\rm G}\) describing the self-gravitating gas that will end up forming prestellar cores:
    \begin{equation}
        \mathcal{P}_\rho(\rho) = B\left(\frac{\rho}{\bar{\rho}}\right)^{-\beta} \ {\rm for} \ \rho\in [\rho_{\rm G}, \rho_{\rm max}].
    \end{equation}
    In the following, we always assume that \(\beta\neq\{2, 3\}\).
    We introduce the parameter \(\epsilon\) that describes the mass fraction of the gas contained in this self-gravitating part of the PDF.
    \begin{align}
        \epsilon &= \int_{\rho_{\rm G}}^{\rho_{\rm max}} B\left(\frac{\rho}{\bar{\rho}}\right)^{-\beta}\frac{\rho}{\bar{\rho}}\frac{{\rm d}\rho}{\bar{\rho}} \\
        &=\frac{B}{2-\beta}\left(\left(\frac{\rho_{\rm max}}{\bar{\rho}}\right)^{2-\beta} - \left(\frac{\rho_{\rm G}}{\bar{\rho}}\right)^{2-\beta}\right),
    \end{align}
    where \(\rho_{\rm max}\) is the maximal density reached in the structures.
    
    Assuming, as shown in Sect. \ref{Sec_density_PS}, that the main contribution to the variance is set by the self-gravitating part, it is given by
    \begin{align}
        {\rm Var}\left(\frac{\rho}{\bar{\rho}}\right) &= \int_{\rho_{\rm G}}^{\rho_{\rm max}} B\left(\frac{\rho}{\bar{\rho}}\right)^{-\beta}\frac{\rho^2}{\bar{\rho^2}}\frac{{\rm d}\rho}{\bar{\rho}} \\
        &=\frac{B}{3-\beta}\left(\left(\frac{\rho_{\rm max}}{\bar{\rho}}\right)^{3-\beta} - \left(\frac{\rho_{\rm G}}{\bar{\rho}}\right)^{3-\beta}\right) \\
        &=\epsilon\frac{2-\beta}{3-\beta}\frac{\left(\frac{\rho_{\rm max}}{\bar{\rho}}\right)^{3-\beta} - \left(\frac{\rho_{\rm G}}{\bar{\rho}}\right)^{3-\beta}}{\left(\frac{\rho_{\rm max}}{\bar{\rho}}\right)^{2-\beta} - \left(\frac{\rho_{\rm G}}{\bar{\rho}}\right)^{2-\beta}}.
    \end{align}
    
    Depending on the value of \(\beta\), we now distinguish three cases and compute the variance in the relevant limit \(\rho_{\rm max}\gg \rho_{\rm G}\):
    \begin{itemize}
        \item First case: \(\beta>3\) (steeper than self-gravity)
        \begin{equation}
            {\rm Var}\left(\frac{\rho}{\bar{\rho}}\right)\simeq\epsilon\frac{\beta-2}{\beta-3}\frac{\rho_{\rm G}}{\bar{\rho}}.
        \end{equation}
        \item Second case: \(3>\beta>2\) (Consistent with self-gravity)
        \begin{equation}
            \label{Eq_PDF_beta_2_3}
            {\rm Var}\left(\frac{\rho}{\bar{\rho}}\right)\simeq\epsilon\frac{\beta-2}{3-\beta}\left(\frac{\rho_{\rm G}}{\bar{\rho}}\right)^{\beta-2}\left(\frac{\rho_{\rm max}}{\bar{\rho}}\right)^{3-\beta}.
        \end{equation}
        \item Third case: \(2>\beta\) (some process  slows down the collapse to enable the accumulation of gas: thermodynamics, magnetic field, rotation)
        \begin{equation}
            {\rm Var}\left(\frac{\rho}{\bar{\rho}}\right)\simeq\epsilon\frac{2-\beta}{3-\beta}\frac{\rho_{\rm max}}{\bar{\rho}}.
        \end{equation}
    \end{itemize}
    
    Because \(\beta=2.5\), we will consider only the second case in this paper.

    \section{The molecular gas mass fraction}
    \label{App_CNM_mass_fraction}
    
    To estimate the molecular gas mass fraction, we rely on a model for the density power spectrum describing the state of the total (neutral+molecular) gas prior to the onset of gravitational collapse. In this regime, where the statistics of the density field is dominated by turbulence, the power spectrum follows a single power law from the injection scale down to the dissipation scale $L_{\rm diss}$:
    \begin{equation}
        \label{Eq_functional_form_PS_no_grav}
        P_{\frac{\rho}{\bar{\rho}}}(k) = {\rm Var}\!\left(\frac{\rho}{\bar{\rho}}\right)
        \begin{cases}
            A, & \dfrac{kL_{\rm inj}}{2\pi} < 1, \\[0.2cm]
            A \left(\dfrac{k L_{\rm inj}}{2\pi}\right)^{-3},
            & 1 < \dfrac{kL_{\rm inj}}{2\pi} < \dfrac{L_{\rm inj}}{L_{\rm cloud}}, \\[0.2cm]
            A \left(\dfrac{k L_{\rm inj}}{2\pi}\right)^{-3},
            & \dfrac{L_{\rm inj}}{L_{\rm cloud}} < \dfrac{kL_{\rm inj}}{2\pi} < \dfrac{L_{\rm inj}}{L_{\rm diss}}.
        \end{cases}
    \end{equation}
    As shown in Sect.~\ref{Sec_density_PS}, the corresponding density variance reads
    \begin{equation}
        \label{Eq_Var_PS_no_grav}
        {\rm Var}\!\left(\frac{\rho}{\bar{\rho}}\right)
        = 32\pi\alpha\left[\frac{1}{3} + \ln\!\left(\frac{L_{\rm inj}}{L_{\rm cloud}}\right)
        + \ln\!\left(\frac{L_{\rm cloud}}{L_{\rm diss}}\right)\right].
    \end{equation}
    The temporal invariance of \(\alpha=M_{\rm inv}/M_{\rm inj}\) implies that the same coefficient appears both in Eq.~\ref{Eq_Var_PS_no_grav}, which describes the turbulence-dominated regime, and in Eq.~\ref{Eq_Var_PS_tot}, which applies once gravity dominates the statistics of the PDF.
    
    We now introduce the variance \(\sigma_{\rm cloud}^2\) of the molecular component assumed to be all contained in clouds and considered as an isolated system. Based on ergodicity, it can be written as:
    \begin{equation}
        \sigma_{\rm cloud}^2 \simeq \frac{1}{V_{\rm cloud}}\int_{V_{\rm cloud}} \rho^2 {\rm d}\tau. 
    \end{equation}
    where \(V_{\rm cloud}\) is the volume of all the molecular gas and \({\rm d}\tau\) is an infinitesimal volume.
    We assume that \(\mathbb{E}(\rho^2)_{\rm cloud}\gg\mathbb{E}(\rho)^2_{\rm cloud}\) given that the high Mach numbers imply for the Larsonian molecular clouds \(\mathbb{E}(\rho)_{\rm cloud}^2/\bar{\rho}_{\rm cloud}^2\simeq (b\mathcal{M})^2\gg 1\). Similarly, the variance of the molecular component within the full system can be written as: 
    \begin{align}
        {\rm Var}(\rho_{\rm cloud})
        &= 32\pi\alpha \bar{\rho}^2 \ln\!\left(\frac{L_{\rm cloud}}{L_{\rm diss}}\right) \\
        &\simeq \frac{1}{V_{\rm tot}}\int_{V_{\rm cloud}} \rho^2 {\rm d}\tau \\
         &\simeq \sigma_{\rm cloud}^2 \epsilon_{\rm cloud}^{\rm V},
    \end{align}
    where \(V_{\rm tot}\) is the volume of the full system, neutral and molecular components and \(\epsilon_{\rm cloud}^{\rm V}=V_{\rm cloud}/V_{\rm tot}\). We thus deduce that the volume fraction is
    \begin{equation}
        \epsilon_{\rm cloud}^{\rm V}
        \simeq \frac{32\pi\alpha\bar{\rho}^2}{\sigma_{\rm cloud}^2}
        \ln\!\left(\frac{L_{\rm cloud}}{L_{\rm diss}}\right),
    \end{equation}
    and the corresponding mass fraction
    \begin{equation}
        \epsilon_{\rm cloud}
        \simeq \frac{32\pi\alpha\bar{\rho}\bar{\rho}_{\rm cloud}}{\sigma_{\rm cloud}^2}
        \ln\!\left(\frac{L_{\rm cloud}}{L_{\rm diss}}\right).
    \end{equation}
    
    This relation does not allow a first-principles prediction of the mass fraction of molecular gas because the coefficient $\alpha$ is not known: it is unclear whether the same $\alpha$ applies to the multiphase ISM and to idealised turbulent boxes since we do not know which mechanism sets it. Although this description neglects the thermal structure of the ISM, it is consistent with the observed $k^{-3}$ density power spectrum and its apparent continuity between the two phases, and therefore provides a reasonable first-order description of the two-phase ISM.

\end{appendix}

\end{document}